\documentclass[conference]{IEEEtran}
\usepackage[latin9]{inputenc}
\usepackage{float}
\usepackage{amsmath}
\usepackage{amssymb}
\usepackage{stmaryrd}
\usepackage{graphicx}
\usepackage{amsthm}
\usepackage{array}

\makeatletter

\floatstyle{ruled}
\newfloat{algorithm}{tbp}{loa}
\providecommand{\algorithmname}{Algorithm}
\floatname{algorithm}{\protect\algorithmname}

\IEEEoverridecommandlockouts
\usepackage{cite}
\usepackage{amsfonts}
\usepackage{textcomp}
\usepackage{xcolor}
\usepackage{epstopdf}

\def\BibTeX{{\rm B\kern-.05em{\sc i\kern-.025em b}\kern-.08em
    T\kern-.1667em\lower.7ex\hbox{E}\kern-.125emX}}

\theoremstyle{plain}

\theoremstyle{remark}

\theoremstyle{plain}

\theoremstyle{plain}

\usepackage[english]{babel}
\providecommand{\corollaryname}{Corollary}
\providecommand{\remarkname}{Remark}
\providecommand{\theoremname}{Theorem}
\providecommand{\lemmaname}{Lemma}

\makeatother

\begin{document}

\title{Joint Estimation of Multipath Angles and Delays for Millimeter-Wave Cylindrical Arrays with Hybrid Front-ends }

\author{Zhipeng Lin, \emph{Member, IEEE}, Tiejun Lv, \emph{Senior Member, IEEE}, Wei Ni, \emph{Senior
Member, IEEE}, J. Andrew Zhang,\\
 \emph{Senior Member, IEEE}, Jie Zeng, \emph{Senior Member, IEEE},
and Ren Ping Liu, \emph{Senior Member, IEEE}

\thanks{Manuscript received October 4, 2019; revised May 22, 2020 and October 30,
2020; accepted February 11, 2021.
\emph{(Corresponding author: Tiejun Lv.)}

Z. Lin, T. Lv and J. Zeng are with the School of Information and Communication
Engineering, BUPT, Beijing, China (e-mail: \{linlzp, lvtiejun\}@bupt.edu.cn;
zengjie@tsinghua.edu.cn). Z. Lin and J. Zeng are also with the School
of Electrical and Data Engineering, UTS, Sydney, Australia.

W. Ni is with the Data61, CSIRO, Sydney, Australia (e-mail: Wei.Ni@data61.csiro.au).

J. A. Zhang and R. P. Liu are with the School of Electrical and Data
Engineering, UTS, Sydney, Australia (e-mail: \{Andrew.Zhang, RenPing.Liu\}@uts.edu.au).}}
\maketitle
\begin{abstract}
Accurate channel parameter estimation is challenging for wideband
millimeter-wave (mmWave) large-scale hybrid arrays, due to beam squint
and much fewer radio frequency (RF) chains than antennas. This paper
presents a novel joint delay and angle estimation approach for wideband
mmWave fully-connected hybrid uniform cylindrical arrays. We first
design a new hybrid beamformer to reduce the dimension of received
signals on the horizontal plane by exploiting the convergence of the
Bessel function, and to reduce the active beams in the vertical direction
through preselection. The important recurrence relationship of the
received signals needed for subspace-based angle and delay estimation
is preserved, even with substantially fewer RF chains than antennas.
Then, linear interpolation is generalized to reconstruct the received
signals of the hybrid beamformer, so that the signals can be coherently
combined across the whole band to suppress the beam squint. As a result,
efficient subspace-based algorithm algorithms can be developed to
estimate the angles and delays of multipath components. The estimated
delays and angles are further matched and correctly associated with
different paths in the presence of non-negligible noises, by putting
forth perturbation operations. Simulations show that the proposed
approach can approach the  Cram\'{e}r-Rao lower bound (CRLB) of the
estimation with a significantly lower computational complexity than
existing techniques.
\end{abstract}


\section{Introduction\label{sec:Introduction}}

\global\long\def\figurename{Fig.}
 Millimeter-wave (mmWave) large-scale antenna arrays, standardized
for the fifth-generation (5G) communication networks, have the potential
to estimate channel parameters with unprecedented accuracy, due to
the excellent directivity of large antenna arrays and the high temporal
resolution provided by mmWave systems \cite{[1],we11,Shahmansoori}.
Accurate channel parameter information plays an important role in
mmWave systems for forming beams with fine accuracy, combating severe
signal attenuation, and suppressing inter-user interference \cite{[4],GC}.
Most existing techniques, such as \cite{Shahmansoori,GC} and \cite{Re2LilongDai},
are only suitable for uniform linear arrays (ULAs) and uniform rectangular
arrays (URAs), whose array steering vectors have linear recurrence
relations. The techniques cannot be directly applied to arrays with
circular layouts, e.g., uniform circular arrays (UCAs) and uniform
cylindrical arrays (UCyAs), as nonlinear recurrence relations exist
between the array steering vectors. However, compared to linear and
rectangular arrays, circular arrays are more compact, have stronger
immunity to mutual coupling, and have stronger immunity to mutual
coupling. They can also provide 360 degrees of angular coverage on
the azimuth plane \cite{Circular1,Circular2}.

Channel parameter estimation techniques have been well studied in
mmWave systems, but limited results are available for large-scale
mmWave antenna arrays using hybrid front-end \cite{andrew,ya2,lin2020tcom}.
A key challenge is that conventional channel parameter estimation
algorithms are inapplicable in mmWave hybrid arrays. Current hybrid
beamforming schemes, typically based on compressed sensing (CS) techniques,
need to discretize channel coefficients and would suffer from accuracy
losses \cite{ya1,hybridESPRIT}. The state-of-the-art spatial spectrum
estimation algorithms, such as maximum likelihood (ML) estimators
\cite{17} and subspace-based algorithms \cite{we11,38,37,34}, were
designed to estimate continuous channel parameters using digital arrays,
where each baseband observation is directly sampled from the signal
received at an antenna. In particular, subspace-based algorithms,
e.g., generalized beamspace method (GBM) \cite{we11}, multiple signal
classification (MUSIC) \cite{38}, estimation of signal parameters
via rotational invariance techniques (ESPRIT) \cite{37}, and quadric
rotational invariance property-based method (QRIPM) \cite{34}, capitalize
on a multiple-invariance structure \cite{strcutrure} of array response
vectors to estimate the channel parameters accurately with dramatically
lower complexities than the ML estimator. The structure exists in
digital arrays, as the received signal of every antenna is available
at the baseband. With a hybrid front-end, the received signals of
multiple antennas are combined via a radio frequency (RF) phase-shifting
network. The multiple-invariance  structure is often obscured or even
lost, and the subspace-based algorithms cannot directly apply.

Challenges also arise from \textit{beam squint} \cite{wideband},
due to typically wide bandwidths of mmWave signals; in other words,
the beam directions can change markedly over the different frequencies
of a signal bandwidth. The beam squint can lead to channel dispersion
in a spatial angle across the bandwidth \cite{wideband}. Most existing
channel parameter estimation methods, e.g., tensor-based subspace
angle estimation (TSAE) \cite{lin2020jsac} and Quasi-Maximum-Likelihood
estimator (Q-MLE) \cite{QMLE}, were designed for narrowband signals,
and hence, do not address the beam squint. One existing solution which
does support wideband operations is incoherent signal-subspace processing
(ISSP) \cite{narrow}. It divides a wide band into non-overlapping
narrow bands. By assuming consistent channel parameters within each
narrowband, channel parameter estimation and localization are applied
repeatedly to the narrow bands, including forming focusing matrices.
Extra steps are also required to combine the results of all the narrow
bands \cite{Inter1}. The complexity of the solution is high.
\begin{figure*}
\begin{centering}
\includegraphics[width=16cm]{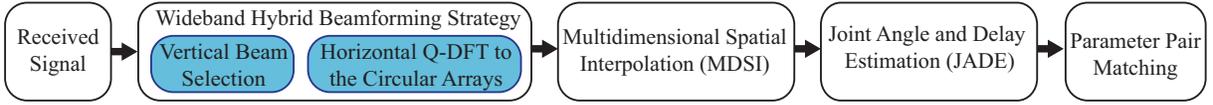}
\par\end{centering}
\centering{}\caption{The flow diagram of the proposed delay and angle estimation approach.\label{fig:The-flow-diagram}}
\end{figure*}

In this paper, we propose a novel joint delay and angle estimation
approach, which enables a hybrid UCyA to estimate the delay and the
azimuth and elevation angles-of-arrival (AOAs) of every impinging
path. Different from any existing works using (typically digital,
narrowband) cylindrical arrays for angle estimation, such as \cite{34,UCYA,we***,UCYA2},
our approach is designed for wideband mmWave hybrid antenna arrays,
addressing the problem of beam squint and requiring far fewer RF chains
than antennas. As depicted in Fig. \ref{fig:The-flow-diagram}, a
series of novel steps are developed in the proposed approach with
the following key contributions.
\begin{itemize}
\item We propose a novel three-dimensional (3D) hybrid beamformer to reduce
the number of required RF chains while preserving the multiple-invariance
structure in array response vectors. As a result, subspace-based algorithms
remain effective for parameter estimation. Specifically, we first
form a small number of vertical beams to pick up significant energy
of received signals. The quasi-discrete Fourier transform (Q-DFT)\footnote{Different from DFT which converts a finite sequence of equally-spaced samples into a
sequence of the same length, Q-DFT can transform the samples to a sequence of a different
length \cite{QDFT}.}, is then conducted on the horizontal plane to convert the received
signals to a small dimension by exploiting the convergence of the
Bessel function.
\item We generalize linear interpolation to the 3D space, to reconstruct
the output signals of the hybrid beamformer. By this means, we achieve
consistent array responses across the wideband and suppress the beam
squint effect. The wideband signals can be coherently combined, and
the high temporal resolution offered by wideband mmWave systems can
be utilized to improve the delay estimation accuracy.
\item We jointly estimate the delay and AOAs of each path, and match the
estimated parameters for different paths. Specifically, the elevation
AOAs and delays are estimated by utilizing ESPRIT to exploit the multiple-invariance
structure, followed by the azimuth AOAs estimated by using MUSIC.
 Perturbation matrices are introduced to mitigate the mismatch between
the estimated delays and angles in the presence of non-negligible
noises. As a result, different paths can be correctly detected.
\end{itemize}

The rest of this paper is organized as follows. The system model is
presented in Section \ref{sec:System-Model}. In Section \ref{sec:Proposed-Two-Step-Wideband},
we develop the two-step hybrid beamforming strategy. The proposed
wideband channel parameter estimation approach is introduced and analyzed
in Section \ref{sec:Proposed-Wideband-3D}. In Section \ref{sec:Simulation-Results},
simulation results are provided to illustrate the performance improvements
of the approach. Finally, conclusions are drawn in Section \ref{sec:Conclusion}.

\textbf{Notation}: $a$, $\mathbf{a}$ and $\mathbf{A}$ stand for
scalar, column vector, and matrix, respectively; $\mathbf{I}_{K}$
represents a $K\times K$ identity matrix, and $\mathbf{0}_{M\times K}$
represents an $M\times K$ zero matrix; $\mathbf{1}_{K}$ denotes
a $K\times1$vector of ones; $[\mathbf{A}]_{i,j}$ is the $(i,j)$-th
entry of $\mathbf{A}$; $\left[\mathbf{A}\right]_{i,:}$ denotes the
$i$-th row of $\mathbf{A}$; the inverse, transpose and conjugate
transpose of $\mathbf{A}$ are $\mathbf{A}^{-1}$, $\mathbf{A}^{T}$
and $\mathbf{A}^{H}$, respectively; $\left\Vert \mathbf{A}\right\Vert _{\textrm{F}}$
and $\textrm{vec}(\mathbf{A})$ denote the Frobenius norm and vectorization
of $\mathbf{A}$, respectively; $\otimes$, $\oplus$ and $\diamond$
denote the Kronecker product, Kronecker sum, and Khatri\textendash Rao
product, respectively; the expectation of a random variable is denoted
by $\mathbb{E}\left\{ \cdot\right\} $; and $O(\cdot)$ denotes the
computational complexity.

\section{System Model \label{sec:System-Model}}

We consider a mmWave multi-antenna orthogonal frequency division multiplexing
(OFDM) system, where a base station (BS) with $N_{\textrm{R}}$ antennas
receives signals from a mobile station (MS)\footnote{An omnidirectional antenna is deployed at the MSs
to maintain connectivity irrespective of the orientation and posture
of the MSs. One of the antenna elements at the BS is set to be the
reference, so that the estimation of the MS would not rotate with
respect to the BS. In the case where a directional antenna is installed
at the MSs, the received signal-to-noise ratio (SNR) at BS could increase
if the BS is inside the mainlobes of the MSs, or decrease otherwise.
This could affect the accuracy of the proposed method in either way,
while the operation of the method is unchanged. }. We assume that the directions and delays of the paths remain unchanged
during parameter estimation. The received signal at subcarrier $m$
$(m=0,1,\ldots,M-1)$ is given by \cite{Shahmansoori}
\begin{equation}
\mathbf{r}_{m}=\mathbf{H}_{m}x_{m}+\mathbf{n}_{m},
\end{equation}
where $\mathbf{H}_{m}\in\mathbb{C}^{N_{\textrm{R}}\times1}$, $\mathbf{n}_{m}\in\mathbb{C}^{N_{\textrm{R}}\times1}$,
and $x_{m}$ denote the channel matrix, the Gaussian noise, and the
transmitted signal for subcarrier $m$, respectively; and $N_{\textrm{p}}$
is the number of paths. The channel matrix, $\mathbf{H}_{m}$, can
be expressed as
\begin{equation}
\mathbf{H}_{m}=\sum_{l=1}^{N_{\textrm{p}}}\beta_{l}e^{-j2\pi f_{m}\tau_{l}}\mathbf{a}_{m}(\phi_{\mathrm{\textrm{R}},l},\theta_{\mathrm{\textrm{R}},l}),\label{eq:H}
\end{equation}
where $\beta_{l}$ is the complex amplitude of the $l$-th path; $\mathbf{a}_{m}(\phi_{\mathrm{\textrm{R}},l},\theta_{\mathrm{\textrm{R}},l})$
is the array response vector with $\phi_{\mathrm{\textrm{R}},l}$
and $\theta_{\mathrm{\textrm{R}},l}$ being the azimuth and elevation
AOAs of the $l$-th path. $\tau_{l}$ is the time delay of the $l$-th
path. $f_{m}$ is the frequency at the $m$-th subcarrier. $f_{m}=f_{0}+m\Delta_{\textrm{F}}$,
where $f_{0}$ is the carrier frequency at the lower end of the band
and $\Delta_{\textrm{F}}$ is the subcarrier spacing. If the signal
bandwidth is much smaller than the carrier frequency, then $f_{m}\approx f_{0}$
and \eqref{eq:H} reverts to the standard narrowband channel model.

\begin{figure*}
\begin{centering}
\includegraphics[width=15cm]{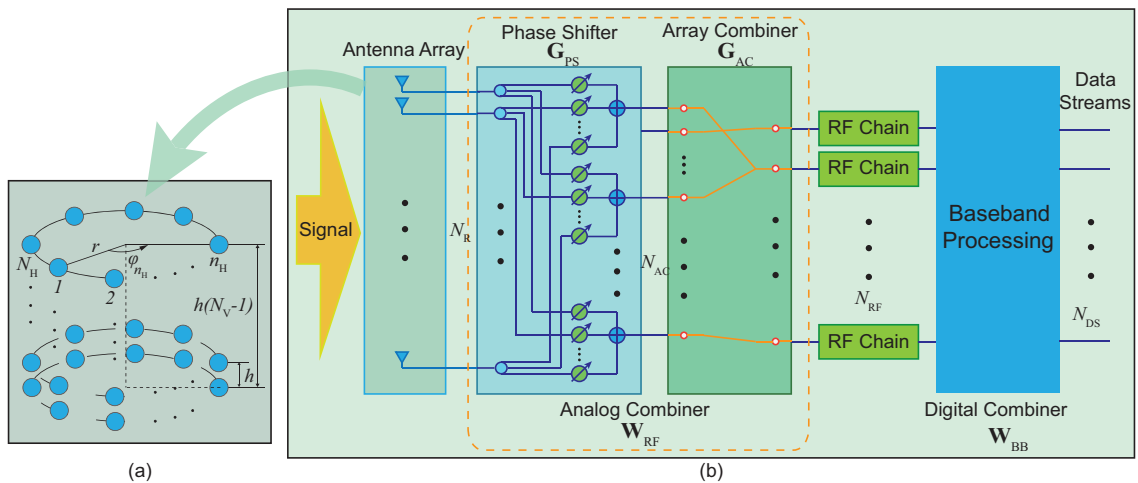}
\par\end{centering}
\centering{}\caption{(a) The geometric model of the UCyA; (b) The block diagram of hybrid
beamforming architecture.\label{fig:The-block-diagram}}
\end{figure*}

The BS uses a hybrid UCyA antenna array. It consists of $N_{\textrm{V}}$
horizontal layers of UCAs, each having $N_{\textrm{H}}$ antennas,
i.e., $N_{\textrm{R}}=N_{\textrm{V}}N_{\textrm{H}}$. The radius of
each UCA is $r$. The vertical distance between any two adjacent UCAs
is $h$. Therefore, the array response vector is given by
\begin{equation}
\mathbf{a}_{m}(\phi_{\mathrm{\textrm{R}},l},\theta_{\mathrm{\textrm{R}},l})=\mathbf{a}_{\textrm{V},m}(\theta_{\mathrm{\textrm{R}},l})\otimes\mathbf{a}_{\textrm{H},m}(\phi_{\mathrm{\textrm{R}},l},\theta_{\mathrm{\textrm{R}},l}),
\end{equation}
where
\begin{align*}
 & [\mathbf{a}_{\textrm{V},m}(\theta_{\mathrm{\textrm{R}},l})]_{n_{\textrm{V}},1}\\
 & =\frac{1}{\sqrt{N_{\textrm{V}}}}\exp\left(-j\frac{2\pi}{c}f_{m}h(n_{\textrm{V}}-\frac{N_{\textrm{V}}+1}{2})\cos(\theta_{\textrm{R},l})\right)
\end{align*}
and
\begin{align*}
 & [\mathbf{a}_{\textrm{H},m}(\phi_{\mathrm{\textrm{R}},l},\theta_{\mathrm{\textrm{R}},l})]_{n_{\textrm{H}},1}\\
 & =\frac{1}{\sqrt{N_{\textrm{H}}}}\exp\left(j\frac{2\pi}{c}f_{m}r\sin(\theta_{\textrm{R},l})\cos(\phi_{\textrm{R},l}-\varphi_{n_{\textrm{H}}})\right)
\end{align*}
are the array response vectors on the vertical and horizontal planes,
respectively, with $n_{\textrm{V}}=1,2,\ldots,N_{\textrm{V}}$ and
$n_{\textrm{H}}=1,2,\ldots,N_{\textrm{H}}$. $c$ is the speed of
light. Here, $\varphi_{n_{\mathrm{H}}}=2\pi(n_{\mathrm{H}}-1)/N_{\mathrm{H}}$
is the difference between the central angles of the $n_{\mathrm{H}}$-th
antenna and the first antenna of each UCA, as shown in Fig. \ref{fig:The-block-diagram}(a).

We consider a hybrid front-end architecture \cite{Hybridprecoding},
as shown in Fig. \ref{fig:The-block-diagram}(b). By applying a hybrid
beamformer, $\mathbf{W}\in\mathbb{C}^{N_{\textrm{R}}\times N_{\textrm{DS}}}$,
to the received signal, $\mathbf{r}_{m}$, the output signal after
beamforming can be expressed as
\begin{equation}
\mathbf{y}_{m}=\mathbf{W}^{H}\mathbf{r}_{m}=\mathbf{W}^{H}\mathbf{H}_{m}x_{m}+\mathbf{W}^{H}\mathbf{n}_{m},
\end{equation}
where the hybrid beamformer, $\mathbf{W}=\mathbf{W}_{\textrm{RF}}\mathbf{W}_{\textrm{BB}}$,
is composed of an analog combiner, $\mathbf{W}_{\textrm{RF}}\in\mathbb{C}^{N_{\textrm{R}}\times N_{\textrm{RF}}}$,
and a digital combiner, \textbf{$\mathbf{W}_{\textrm{BB}}\in\mathbb{C}^{N_{\textrm{RF}}\times N_{\textrm{DS}}}$}.
$N_{\textrm{RF}}$ and $N_{\textrm{DS}}$ are the numbers of RF chains
and data streams, respectively.

We further divide the analog combiner, $\mathbf{W}_{\textrm{RF}}$,
into an array combiner set, $\mathbf{G}_{\textrm{AC}}\in\mathbb{C}^{N_{\textrm{AC}}\times N_{\textrm{RF}}}$,
and a phase shifter set, $\mathbf{G}_{\textrm{PS}}\in\mathbb{C}^{N_{\textrm{R}}\times N_{\textrm{AS}}}$,
i.e., $\mathbf{W}_{\textrm{RF}}=\mathbf{G}_{\textrm{PS}}\mathbf{G}_{\textrm{AC}}$.
$N_{\textrm{AC}}$ is the number of the combiners deployed in the
array combiner set. $N_{\textrm{R}}\geq N_{\textrm{AC}}\geq N_{\textrm{RF}}\geq N_{\textrm{DS}}$.
As illustrated in Fig. \ref{fig:The-block-diagram}, $\mathbf{G}_{\textrm{PS}}$
is a phase shifter matrix with elements given by $\left[\mathbf{G}_{\textrm{PS}}\right]_{n_{\textrm{R}},n{}_{\textrm{AS}}}=\exp(j\xi)$
$(\xi\in\mathbb{R},\:n_{\textrm{R}}=1,2,\ldots,N_{\textrm{R}},$ and
$n_{\textrm{AC}}=1,2,\ldots,N_{\textrm{AC}})$. $\mathbf{G}_{\textrm{AC}}$
is a binary matrix, and its entry $\left[\mathbf{G}_{\textrm{AC}}\right]_{n_{\textrm{AC}},n_{\textrm{RF}}}\in\left\{ 0,1\right\} $$(n_{\textrm{RF}}=1,2,\ldots,N_{\textrm{RF}})$.
Here, the role of $\mathbf{W}_{\textrm{BB}}$ is to guarantee the
power constraint.

\section{Proposed Two-Step Wideband Hybrid Beamforming Strategy\label{sec:Proposed-Two-Step-Wideband}}

In this section, new hybrid beamformers are designed to select the
meaningful beams needed vertically for angle and delay estimation,
and transform the received high-dimensional signals of horizontal
UCAs to be low-dimensional by taking Q-DFT and the convergence property
of the Bessel function. We prove that the number of low dimensions
does not grow with the number of antennas per UCA. The minimal number
of required RF chains is the product of the number of vertical beams
and the number of low dimensions.

It is worth mentioning that all the beamformers we design here are
linear transforms. Therefore, the critical invariance structure for
the validity of ESPRIT for the angle and delay estimation, can be
recovered losslessly between respective submatrices of the space-time
response matrix for the subsequent angle and delay estimation.

\subsection{Step 1: Vertical Beam Selection\label{subsec:Step-1:-Vertical}}

We first propose a new hybrid beamformer, denoted by $\mathbf{W}_{\textrm{s1}}$,
in the vertical beamspace. By exploiting the sparsity (or low rank)
nature of mmWave multi-antenna channels, the vertical beams can be
selected: i) to estimate the number of paths; and ii) to determine
the number of vertical beams needed for the joint angle and delay estimation (JADE)
(to be developed in Section \ref{subsec:Wideband-JADE-Algorithm}).

The output signal after the vertical beamforming is $\mathbf{y}_{\textrm{s1,}m}=\mathbf{W}_{\textrm{s1}}^{H}\mathbf{r}_{m}\in\mathbb{C}^{N_{\textrm{DS},1}\times1}.$
The hybrid beamformer conducts vertical beamspace transforming and
can be constructed as $\mathbf{W}_{\textrm{s1}}=\mathbf{G}_{\textrm{PS},\textrm{s1}}\mathbf{G}_{\textrm{AC},\textrm{s1}}\mathbf{W}_{\textrm{BB},\textrm{s1}},$
where $\mathbf{W}_{\textrm{BB},\textrm{s1}}=\frac{1}{\sqrt{N_{\textrm{V}}}}\mathbf{I}_{N_{\textrm{V}}}\in\mathbb{C}^{N_{\textrm{V}}\times N_{\textrm{V}}}$,
$\mathbf{G}_{\textrm{PS},\textrm{s1}}=\mathbf{U}_{\textrm{d}}\otimes\mathbf{I}_{N_{\textrm{H}}}\in\mathbb{C}^{N_{\textrm{R}}\times N_{\textrm{R}}}$,
and $\mathbf{G}_{\textrm{AC},\textrm{s1}}=\left[\mathbf{I}_{N_{\textrm{V}}}\otimes\mathbf{1}_{N_{\textrm{H}}}^{T}\right]^{T}\in\mathbb{C}^{N_{\textrm{R}}\times N_{\textrm{V}}}.$
Here, $\mathbf{U}_{\textrm{d}}$ contains $N_{\textrm{V}}$ orthogonal
array response vectors corresponding to $N_{\textrm{V}}$ vertically,
angularly evenly spaced beams. $\mathbf{U}_{\textrm{d}}=\left[\mathbf{U}_{\textrm{d},1},\mathbf{U}_{\textrm{d},2},\ldots,\mathbf{U}_{\textrm{d},N_{\textrm{V}}}\right]\in\mathbb{C}^{N_{\textrm{V}}\times N_{\textrm{V}}},$
where
\begin{align}
\mathbf{U}_{\textrm{d},i} & =[\exp(-j\frac{2\pi}{N_{\textrm{V}}}(-\frac{N_{\textrm{V}}-1}{2})i),\exp(-j\frac{2\pi}{N_{\textrm{V}}}(-\frac{N_{\textrm{V}}-3}{2})i),\nonumber \\
 & \ldots,\exp(-j\frac{2\pi}{N_{\textrm{V}}}(\frac{N_{\textrm{V}}-1}{2})i)]^{T},\,i=1,2,\ldots,N_{\textrm{V}}.\label{eq:Ud}
\end{align}
Thus at this step, the numbers of both data streams and RF chains
are equal to that of beams, i.e., $N_{\textrm{DS},1}=N_{\textrm{RF},1}=N_{\textrm{V}}$.
The number of array combiners is equal to that of receive antennas,
i.e., $N_{\textrm{AC},1}=N_{\textrm{R}}$. The $i$-th element of
$\mathbf{y}_{\textrm{s1,}m}$ can be written as
\begin{align}
\left[\mathbf{y}_{\textrm{s1},m}\right]_{i,1} & =\left[\left(\mathbf{G}_{\textrm{PS},\textrm{s1}}\mathbf{G}_{\textrm{AC},\textrm{s1}}\mathbf{W}_{\textrm{BB},\textrm{s1}}\right)^{H}\mathbf{r}_{m}\right]_{i}\nonumber \\
 & =\frac{1}{\sqrt{N_{\textrm{V}}}}\mathbf{U}_{\textrm{d},i}^{H}(\mathbf{I}_{N_{\textrm{V}}}\otimes\mathbf{1}_{N_{\textrm{H}}}^{T})\mathbf{r}_{m}.
\end{align}
The total beam power at the $m$-th subcarrier is given by
\begin{equation}
\sigma_{m}^{2}=\mathbf{y}_{\textrm{s1},m}^{H}\mathbf{y}_{\textrm{s1,}m}=\sum_{i=1}^{N_{\textrm{V}}}\sigma_{m,i}^{2},
\end{equation}
where $\sigma_{m,i}^{2}=\left|\left[\mathbf{y}_{\textrm{s1},m}\right]_{i,1}\right|^{2}$
is the power of the $i$-th beam which depends on the AOA of the impinging
signal inside the beam. Given the sparsity of mmWave multi-antenna
channels, the signal power is concentrated in a small number of beams.
We select the dominant beams at the $m$-th subcarrier by defining
an index selection set $\mathcal{U}_{m}$, as given by
\begin{equation}
\mathcal{U}_{m}\triangleq\left\{ \eta(1),\eta(2),\ldots,\eta(N_{\textrm{B},m})\right\} ,\label{eq:15new}
\end{equation}
where $N_{\textrm{B},m}$ is the number of selected beams, and $\eta(u_{m})$
is the index for $\sigma_{m,\eta(u_{m})}^{2}$ with $u_{m}=1,2,\ldots,N_{\textrm{B},m}$.
$\eta(u_{m})$ can be obtained as
\[
\begin{cases}
\eta(1)=\arg\max_{i\in\{1,\ldots,N_{\textrm{V}}\}}\sigma_{m,i}^{2},\\
\eta(2)=\arg\max_{i\in\{1,\ldots,N_{\textrm{V}}\}\setminus\{\eta(1)\}}\sigma_{m,i}^{2},\\
\quad\vdots\\
\eta(N_{\textrm{B},m})=\arg\max_{i\in\{1,\ldots,N_{\textrm{V}}\}\setminus\{\eta(1),\ldots,\eta(N_{\textrm{B},m}-1)\}}\sigma_{m,i}^{2}.
\end{cases}
\]
The following criterion can be used to decide $N_{\textrm{B},m}$
and select the $N_{\textrm{B},m}$ strongest beams:
\begin{equation}
\sum_{u_{m}=1}^{N_{\textrm{B},m}}\sigma_{m,\eta(u_{m})}^{2}\geq\eta\sigma_{m}^{2},\label{eq:select}
\end{equation}
where $\eta$ is a power threshold which can be empirically specified.
$\eta$ can be selected close to 1, e.g., $\eta=0.9$, as paths reflected
more than once, and diffuse scattering, account for less than 10\%
of the total energy, as found in \cite{MGP+14}\footnote{It is shown in \cite{MGP+14} that for mmWave systems, the contributions
of paths reflected more than once, and the diffuse scattering components
are weak, only accounting for less than 10\% of the total energy. }. Moreover, mmWave signals fade rapidly when reflecting off a surface
\cite{3DMODEL}, and become barely distinguishable from noises after
two reflections \cite{Olivier,ya2,MGP+14}.

\begin{figure*}
\begin{centering}
\includegraphics[width=14cm]{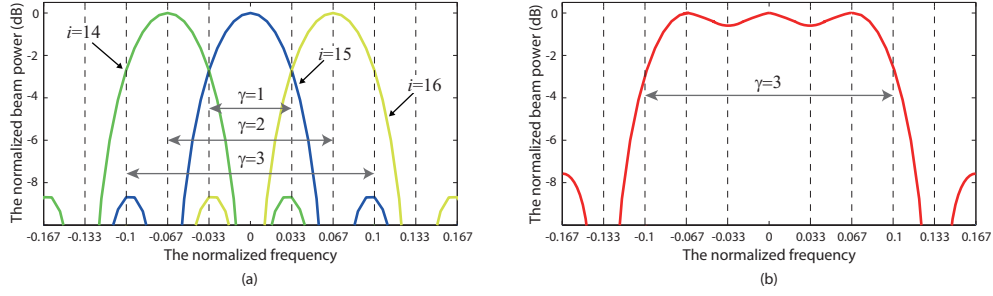}
\par\end{centering}
\centering{}\caption{The normalized beam power as a function of the normalized frequency.
(a) The power of the$14$-th, $15$-th, and $16$-th beams; (b) The
power of the combined three beams.\label{fig:Normalizedpower}}
\end{figure*}

There is dispersion in the angular domain across the bandwidth in
multi-antenna wireless systems \cite{wideband}. We first assume that
the transmission channel at each subcarrier is narrowband. Because
of small dispersion in narrowband systems, the number of orthogonal
beams in the vertical beamspace is equal to the number of received
paths, i.e., $N_{\textrm{B},m}=N_{\textrm{p}}$ \cite{wideband}.
However, the dispersion can have a non-negligible effect in broadband
systems, such as the one considered in this paper, where a point source
spreads across spatial angle and time. A strong dispersion would result
in severe power loss and pulse distortion, if not addressed properly,
and affect the follow-on angle and delay estimation. The dispersion
effect can be characterized by the \textit{channel dispersion factor},
$\gamma$, as specified by \cite{wideband}
\begin{equation}
\gamma=\frac{1}{N_{\textrm{p}}}\sum_{l=1}^{N_{\textrm{p}}}\gamma_{l}=\frac{1}{N_{\textrm{p}}}\sum_{l=1}^{N_{\textrm{p}}}N_{\textrm{V}}\alpha\left|\chi_{\textrm{c},l}\right|,
\end{equation}
where $\alpha=W/f_{\textrm{c}}$ is the fractional bandwidth, $\chi_{\textrm{c},l}=f_{\textrm{c}}h\cos(\theta_{\textrm{R},l})/c$
is the normalized beam angle, $W$ is the signal bandwidth, and $f_{\textrm{c}}$
is the center frequency.

To illustrate the impact of the dispersion, we assume that the system
operates at $f_{\textrm{c}}=$30 GHz and the transmitted signal has
unit amplitude. For simplicity, it is assumed that only one path with
$\beta_{l}=1$ and $\theta_{\textrm{R},l}=60^{\circ}$, and the number
of beams is $N_{\textrm{V}}=60$. We have $\chi_{\textrm{c},l}\approx0.25$,
which corresponds to the $i_{0}=15$-th beam. Fig. \ref{fig:Normalizedpower}(a)
shows the normalized power, $\left|\mathbf{y}_{\textrm{s1},i}(f)\right|^{2}/N_{\textrm{V}}$,
of the $14$-th, $15$-th, and $16$-th beams versus the normalized
frequency, $f_{\textrm{normal}}=f/f_{\textrm{c}},$ for $f\in[-W/2,W/2]$\footnote{For convenience, here we only plot the beam power as a function of
continuous frequency for illustration.}. We can see that, if the normalized frequency $f_{\textrm{normal}}<0.033,$
i.e., the channel dispersion factor $\gamma<1$, the beams in Fig.
\ref{fig:Normalizedpower}(a) do not affect one another within the
bandwidth, $W$. If $\gamma>1$, power loss and interference may occur.
To prevent this from happening, the $\gamma$ adjacent beams centered
at $i_{0}$ need to be taken into consideration. In the case of $\gamma=3$,
Fig. \ref{fig:Normalizedpower}(b) plots the normalized power of the
three beams combined, i.e., the $14$-th, $15$-th, and $16$-th beams.
It can be seen that, by combining the three beams, the normalized
power becomes approximately flat across the operating band.

Because of the dispersion, we have to jointly consider $N_{\textrm{B}}=\gamma N_{\textrm{p}}$
vertically spaced beams to include all possible beams, as the normalized
beam angle, $\chi_{\textrm{c},l}$, is unknown. The overall index
selection set $\mathcal{U}$ is given by
\begin{equation}
\mathcal{U}=\mathcal{U}_{0}\cup\mathcal{U}_{1}\cup\ldots\cup\mathcal{U}_{M-1},
\end{equation}
where the elements of $\mathcal{U}$ are $\eta(u)$ with $u=1,2,\ldots,N_{\textrm{B}}.$
It is possible that the same indices are picked up at different subcarriers
because of the dispersion, e.g., $\eta(u_{m})=\eta(u_{m'})$ for $m'\in\{0,\ldots,M-1\}\setminus\{m\}$.
We have $N_{\textrm{B}}=\gamma N_{\textrm{p}}\leq\sum_{m=0}^{M-1}N_{\textrm{B},m}$
to avoid overlooking significant paths in the subsequent joint delay
and angle  estimation process. Algorithm \ref{alg:algorithmNB} summarizes
the procedure of the beam selection at this step\footnote{Card$(\mathcal{U})$ in Algorithm \ref{alg:algorithmNB} denotes the
cardinality of the set $\mathcal{U}$.}.

\begin{algorithm}
\protect\caption{Beam selection algorithm \label{alg:algorithmNB}}

\begin{itemize}
\item \textbf{Input}: The processed signals, $\mathbf{y}_{\textrm{s1,}m}$,$(m=0,1,\ldots,M-1)$,
the beam number, $N_{\textrm{V}}$, and the threshold, $\eta$.
\item \textbf{Output}: The overall selection set, $\mathcal{U}$, and the
estimated number of significant beams, $N_{\textrm{B}}$.
\item \textbf{Initialization}: $\mathcal{U}=\mathcal{U}_{0}=\mathcal{U}_{1}=\ldots=\mathcal{U}_{M-1}=\emptyset$,
\item \textbf{For} $m=0$ \textbf{to} $M-1$ \textbf{do}
\begin{itemize}
\item Set $\mathcal{V}=\emptyset.$
\item \textbf{For} $i=1$ \textbf{to} $N_{\textrm{V}}$ \textbf{do}
\begin{itemize}
\item $\sigma_{m,i}^{2}=\left|\mathbf{y}_{\textrm{s1},m,i}\right|^{2}$,
and update$\mathcal{V}=\mathcal{V}\cup\{\sigma_{m,i}^{2}\}$.
\end{itemize}
\item \textbf{End for}
\item $\sigma_{m}^{2}=\sum_{i=1}^{N_{\textrm{V}}}\sigma_{m,i}^{2}.$
\item \textbf{While} $\sum_{\eta(u_{m})\in\mathcal{U}_{m}}\sigma_{m,\eta(u_{m})}^{2}<\eta\sigma_{m}^{2}$
\textbf{do}
\begin{itemize}
\item Find the largest $\sigma_{m,i}^{2}$ from $\mathcal{V}$, and update
$\mathcal{U}_{m}=\mathcal{U}_{m}\cup\{i\}$ and $\mathcal{V}=\mathcal{V}\setminus\{\sigma_{m,i}^{2}\}$.
\end{itemize}
\item \textbf{End while}
\item Update $\mathcal{U}=\mathcal{U}\cup\mathcal{U}_{m}$.
\end{itemize}
\item \textbf{End for}
\item $N_{\textrm{B}}=$Card$(\mathcal{U})$.
\end{itemize}
\end{algorithm}

\subsection{Step 2: Horizontal Q-DFT Beamforming\label{subsec:Step-2:-Horizontal}}

We proceed to design a new hybrid beamformer to transform the high-dimensional
signals of each UCA to a low dimension requiring much fewer RF chains
than array elements. This is achieved by first applying Q-DFT to the
signals and then exploiting the convergence property of the Bessel
function to remove insignificant dimensions.

We first derive an approximate expression for the array response vector
to explain the design rationale of this step. According to the \textit{Jacobi-Anger
expansion} \cite{bessel}, the $n_{\textrm{H}}$-th array response
vector on the horizontal plane can be written as
\begin{align}
[\mathbf{a}_{\textrm{H},m}(\phi_{\mathrm{\textrm{R}},l},\theta_{\mathrm{\textrm{R}},l})]_{n_{\textrm{H}},1}=\frac{1}{\sqrt{N_{\textrm{H}}}}e^{j\varpi_{m,l}\cos(\phi_{\textrm{R},l}-\varphi_{n_{\textrm{H}}})}\nonumber \\
=\frac{1}{\sqrt{N_{\textrm{H}}}}\sum_{q=-\infty}^{\infty}j^{q}J_{q}(\varpi_{m,l})e^{jq(\phi_{\textrm{R},l}-\varphi_{n_{\textrm{H}}})},\label{eq:multiplier}
\end{align}
where $\varpi_{m,l}=\frac{2\pi}{c}f_{m}r\sin(\theta_{\textrm{R},l})$
and $J_{q}(\varpi_{m,l})$ is the Bessel function of the first kind
of order $q$.

We notice that the last multiplier in \eqref{eq:multiplier}, i.e.,
$e^{jq(\phi_{\textrm{R},l}-\varphi_{n_{\textrm{H}}})}=e^{jq\phi_{\textrm{R},l}-j2\pi q(n_{\mathrm{H}}-1)/N_{\mathrm{H}}},$
is of strong resemblance to the weight vectors in the DFT. We take
the Q-DFT \cite{book11} to transform the horizontal array response
vectors \eqref{eq:multiplier} to offset $\varphi_{n_{\textrm{H}}}$.
The $p$-th order Q-DFT of \eqref{eq:multiplier} can be expressed
as
\begin{align}
 & A_{\textrm{PM},p}=\sum_{n_{\textrm{H}}=1}^{N_{\textrm{H}}}\left([\mathbf{a}_{\textrm{H},m}(\phi_{\mathrm{\textrm{R}},l},\theta_{\mathrm{\textrm{R}},l})]_{n_{\textrm{H}},1}\right)e^{-j\frac{2\pi(n_{\mathrm{H}}-1)}{N_{\mathrm{H}}}p}\nonumber \\
 & =\frac{1}{\sqrt{N_{\textrm{H}}}}\sum_{n_{\textrm{H}}=1}^{N_{\textrm{H}}}\sum_{q=-\infty}^{\infty}j^{q}J_{q}(\varpi_{m,l})e{}^{jq(\phi_{\textrm{R},l}-\varphi_{n_{\textrm{H}}})-jp\varphi_{n_{\textrm{H}}}}\nonumber \\
 & =\frac{1}{\sqrt{N_{\textrm{H}}}}\sum_{n_{\textrm{H}}=1}^{N_{\textrm{H}}}\left(\sum_{q=-\infty}^{\infty}j^{q}J_{q}(\varpi_{m,l})e^{-j\varphi_{n_{\textrm{H}}}(p+q)}e^{jq\phi_{\textrm{R},l}}\right).\label{eq:dft}
\end{align}
Let $p+q=QN_{\mathrm{H}}$, i.e., $q=QN_{\mathrm{H}}-p$. Then, \eqref{eq:dft}
can be rewritten as
\begin{align}
A_{\textrm{PM},p} & =\frac{1}{\sqrt{N_{\textrm{H}}}}\sum_{Q=-\infty}^{\infty}N_{\textrm{H}}j^{(QN_{\mathrm{H}}-p)}\nonumber \\
 & \times J_{(QN_{\mathrm{H}}-p)}(\varpi_{m,l})e^{j(QN_{\mathrm{H}}-p)\phi_{\textrm{R},l}}\nonumber \\
 & \stackrel{(\star)}{=}\sqrt{N_{\textrm{H}}}\left[j^{p}J_{p}(\varpi_{m,l})e^{-jp\phi_{\textrm{R},l}}\vphantom{\sum_{Q=-\infty,Q\neq0}^{\infty}}\right.\nonumber \\
 & \left.+\sum_{Q=-\infty,Q\neq0}^{\infty}\varepsilon_{p,Q}(\varpi_{m,l},\phi_{\textrm{R},l})\right],\label{eq:dft2}
\end{align}
where
\[
\varepsilon_{p,Q}(\varpi_{m,l},\phi_{\textrm{R},l})=j^{(QN_{\mathrm{H}}-p)}J_{(QN_{\mathrm{H}}-p)}(\varpi_{m,l})e^{j(QN_{\mathrm{H}}-p)\phi_{\textrm{R},l}}.
\]
$(\star)$ is obtained by the property of the Bessel function $J_{-v}(x)=(-1)^{v}J_{v}(x)$
\cite{bessel}.

\noindent \textbf{Lemma 1}. \textit{For the Bessel function $J_{v}(x)$,
when its order $v$ is larger than its argument $x$, i.e., $\left|v\right|>\left|x\right|$,
the amplitude of $J_{v}(x)$ is so small and negligible, i.e., $\left|J_{v}(x)\right|\approx0.$}

\noindent \textit{Proof.} See Appendix I.

From Lemma 1, we can derive the following theorem on the approximation
of the horizontal array response vector.

\noindent \textbf{Theorem 1. }\textit{If $N_{\mathrm{H}}\geq2P$,
the $N_{\mathrm{H}}$-dimensional array response vectors on the horizontal
plane can be transformed to a much smaller $(2P+1)$-dimensional space
with negligible loss, i.e., $p\in[-P,P]\cap\mathbb{Z}$, where $P=\left\lfloor 2\pi f_{0}r/c\right\rfloor $
is the highest order. The $p$-th order of the $(2P+1)$-dimensional
vector, $A_{\textrm{PM},p},$ can be approximated as $A_{\textrm{PM},p}\approx\sqrt{N_{\textrm{H}}}j^{p}J_{p}(\varpi_{m,l})\exp\left(-jp\phi_{\textrm{R},l}\right).$}

\noindent \textit{Proof.} See Appendix II.

According to Theorem 1, we see that, by using the Q-DFT, the $N_{\textrm{H}}$-dimensional
array response vectors of the horizontal UCA can be transformed to
only $(2P+1)$ dimensions, and each element of the vector can be approximately
expressed as an exponential function weighted by a Bessel function
of the same order, as long as the conditions in Theorem 1, i.e., $N_{\textrm{H}}\geq2P$,
is met\footnote{In general, this condition can be met in large-scale antenna array
systems, where a large number of antennas are deployed. }.

We note that Q-DFT is a linear transform and hence can preserve the
multiple-invariance structure of the array response vectors for the
subsequent angle and delay estimation, as will be elaborated on in
Section \ref{subsec:Wideband-JADE-Algorithm}; see \eqref{eq:ud}
and \eqref{eq:beamrelation}. Thus, combining with the beams selected
at Step 1, we can design the values of the phase shifters based on
the Q-DFT and the beamspace transform to convert the array response
vectors to a low dimension. Only a small number of RF chains are needed
for the  delay and angle estimation.

At this step, the hybrid beamformer is $\mathbf{W}_{\textrm{s2}}=\mathbf{G}_{\textrm{PS},\textrm{s2}}\mathbf{G}_{\textrm{AC},\textrm{s2}}\mathbf{W}_{\textrm{BB},\textrm{s2}}$.
Then we have
\begin{equation}
\mathbf{y}_{\textrm{s2,}m}=\mathbf{W}_{\textrm{s2}}^{H}\mathbf{r}_{m}=(\mathbf{G}_{\textrm{PS},\textrm{s2}}\mathbf{G}_{\textrm{AC},\textrm{s2}}\mathbf{W}_{\textrm{BB},\textrm{s2}})^{H}\mathbf{r}_{m}\in\mathbb{C}^{N_{\textrm{DS},2}\times1},\label{eq:receivedS2}
\end{equation}
where $\mathbf{W}_{\textrm{BB},\textrm{s2}}=\sqrt{N_{\textrm{V}}/N_{\textrm{H}}}\mathbf{I}_{(2P+1)N_{\textrm{B}}}\in\mathbb{C}^{(2P+1)N_{\textrm{B}}\times(2P+1)N_{\textrm{B}}}$,
$\mathbf{J}_{\textrm{B}}=[\mathbf{J}_{\textrm{B},1},\mathbf{J}_{\textrm{B},2},\ldots,\mathbf{J}_{\textrm{B},N_{\textrm{B}}}]\in\mathbb{R}^{N_{\textrm{V}}\times N_{\textrm{B}}}$,
and $\mathbf{G}_{\textrm{AC},\textrm{s2}}=\mathbf{J}_{\textrm{B}}\otimes\mathbf{I}_{(2P+1)}\in\mathbb{R}^{(2P+1)N_{\textrm{V}}\times(2P+1)N_{\textrm{B}}}$.
The elements of $\mathbf{J}_{\textrm{B},u}\in\mathbb{R}^{N_{\textrm{V}}\times1}$
are given by
\[
\left[\mathbf{J}_{\textrm{B},u}\right]_{n_{\textrm{V}},1}=\begin{cases}
1, & \textrm{if }n_{\textrm{V}}=\eta(u);\\
0, & \textrm{otherwise.}
\end{cases}
\]

We design the phase shifter set of the analog part of the hybrid array,
as $\mathbf{G}_{\textrm{PS},\textrm{s2}}=\mathbf{U}_{\textrm{d}}\otimes\mathbf{U}_{\textrm{sH}}\in\mathbb{C}^{N_{\textrm{R}}\times(2P+1)N_{\textrm{V}}},$
where $\mathbf{U}_{\textrm{d}}$ is given in \eqref{eq:Ud} and the
element of $\mathbf{U}_{\textrm{sH}}\in\mathbb{C}^{N_{\textrm{H}}\times(2P+1)}$
can be expressed as $\left[\mathbf{U}_{\textrm{sH}}\right]_{n_{\mathrm{H}},p+P+1}=e{}^{j2\pi(n_{\mathrm{H}}-1)p/N_{\mathrm{H}}}$.
Hence, the analog combiner of the hybrid array can be constructed
as
\begin{align}
\mathbf{W}_{\textrm{RF},\textrm{s2}} & =\mathbf{G}_{\textrm{PS},\textrm{s2}}\mathbf{G}_{\textrm{AC},\textrm{s2}}=(\mathbf{U}_{\textrm{d}}\otimes\mathbf{U}_{\textrm{sH}})(\mathbf{J}_{\textrm{B}}\otimes\mathbf{I}_{(2P+1)})\nonumber \\
 & \stackrel{(\star)}{=}(\mathbf{U}_{\textrm{d}}\mathbf{J}_{\textrm{B}})\otimes(\mathbf{U}_{\textrm{sH1}}\mathbf{I}_{(2P+1)})=\mathbf{U}_{\textrm{sV}}\otimes\mathbf{U}_{\textrm{sH}},
\end{align}
where $(\star)$ follows a property of the Khatri-Rao product, i.e.,
$(\mathbf{A}\otimes\mathbf{B})(\mathbf{C}\otimes\mathbf{D})=\mathbf{AC}\otimes\mathbf{BD}$.
The element of $\mathbf{U}_{\textrm{sV}}\in\mathbb{C}^{N_{\textrm{V}}\times N_{\textrm{B}}}$
can be calculated as $\left[\mathbf{U}_{\textrm{sV}}\right]_{n_{\textrm{V}},u}=\exp(-j\frac{2\pi}{N_{\textrm{V}}}(\frac{N_{\textrm{V}}+1}{2}-n_{\textrm{V}})\eta(u))$.
As a result, at this step we have $N_{\textrm{DS},2}=N_{\textrm{RF},2}=(2P+1)N_{\textrm{B}}$
data streams and RF chains, and $N_{\textrm{AC,2}}=(2P+1)N_{\textrm{V}}$
array combiners. The processed received signal in \eqref{eq:receivedS2}
can be written as
\begin{align}
\mathbf{y}_{\textrm{s2,}m} & =\mathbf{W}_{\textrm{s2}}^{H}\mathbf{r}_{m}\nonumber \\
 & =(\mathbf{G}_{\textrm{s2}}\mathbf{W}_{\textrm{RF},\textrm{s2}}\mathbf{W}_{\textrm{BB},\textrm{s2}})^{H}\mathbf{H}_{m}x_{m}+\mathbf{W}_{\textrm{s2,}m}^{H}\mathbf{n}_{m}\nonumber \\
 & =\sqrt{\frac{N_{\textrm{V}}}{N_{\textrm{H}}}}(\mathbf{U}_{\textrm{sV}}\otimes\mathbf{U}_{\textrm{sH}})^{H}\sum_{l=1}^{N_{\textrm{p}}}\beta_{l}x_{m}e^{-j2\pi f_{m}\tau_{l}}\nonumber \\
 & \qquad\times\mathbf{a}_{m}(\phi_{\mathrm{\textrm{R}},l},\theta_{\mathrm{\textrm{R}},l})+\mathbf{W}_{\textrm{s2,}m}^{H}\mathbf{n}_{m}\nonumber \\
 & \stackrel{(\star)}{=}\sqrt{\frac{N_{\textrm{V}}}{N_{\textrm{H}}}}\sum_{l=1}^{N_{\textrm{p}}}\beta_{l}x_{m}e^{-j2\pi f_{m}\tau_{l}}(\mathbf{U}_{\textrm{sV}}^{H}\otimes\mathbf{U}_{\textrm{sH}}^{H})\nonumber \\
 & \qquad\times(\mathbf{a}_{\textrm{V},m}(\theta_{\mathrm{\textrm{R}},l})\otimes\mathbf{a}_{\textrm{H},m}(\phi_{\mathrm{\textrm{R}},l},\theta_{\mathrm{\textrm{R}},l}))+\mathbf{W}_{\textrm{s2,}m}^{H}\mathbf{n}_{m}\nonumber \\
 & =\sum_{l=1}^{N_{\textrm{p}}}\beta_{l}x_{m}e^{-j2\pi f_{m}\tau_{l}}(\sqrt{N_{\textrm{V}}}\mathbf{U}_{\textrm{sV}}^{H}\mathbf{a}_{\textrm{V},m}(\theta_{\mathrm{\textrm{R}},l}))\nonumber \\
 & \qquad\otimes(\frac{1}{\sqrt{N_{\textrm{H}}}}\mathbf{U}_{\textrm{sH}}^{H}\mathbf{a}_{\textrm{H},m}(\phi_{\mathrm{\textrm{R}},l},\theta_{\mathrm{\textrm{R}},l}))+\mathbf{W}_{\textrm{s2,}m}^{H}\mathbf{n}_{m}\nonumber \\
 & =\sum_{l=1}^{N_{\textrm{p}}}\beta_{l}x_{m}e^{-j2\pi f_{m}\tau_{l}}(\tilde{\mathbf{a}}_{\textrm{V},m}(\theta_{\mathrm{\textrm{R}},l})\nonumber \\
 & \qquad\otimes\tilde{\mathbf{a}}_{\textrm{H},m}(\phi_{\mathrm{\textrm{R}},l},\theta_{\mathrm{\textrm{R}},l}))+\mathbf{W}_{\textrm{s2,}m}^{H}\mathbf{n}_{m},\label{eq:sp2channel}
\end{align}
where $(\star)$ stems from another property of the Khatri-Rao product,
i.e., $(\mathbf{A}\otimes\mathbf{B})^{H}=\mathbf{A}^{H}\otimes\mathbf{B}^{H}$.
According to Theorem 1, the elements of the resulting vertical and
horizontal array response vectors $\tilde{\mathbf{a}}_{\textrm{V},m}(\theta_{\mathrm{\textrm{R}},l})$
and $\tilde{\mathbf{a}}_{\textrm{H},m}(\phi_{\mathrm{\textrm{R}},l},\theta_{\mathrm{\textrm{R}},l})$
are given by
\begin{align}
 & [\tilde{\mathbf{a}}_{\textrm{V},m}(\theta_{\mathrm{\textrm{R}},l})]_{u,1}=\sqrt{N_{\textrm{V}}}\mathbf{U}_{\textrm{sV}}^{H}\mathbf{a}_{\textrm{V},m}(\theta_{\mathrm{\textrm{R}},l})\nonumber \\
 & \qquad=\sum_{n_{\textrm{V}}=1}^{N_{\textrm{V}}}\exp\left(-j\frac{2\pi}{c}f_{m}h(n_{\textrm{V}}-\frac{N_{\textrm{V}}+1}{2})\cos(\theta_{\textrm{R},l})\right)\nonumber \\
 & \qquad\qquad\times\exp\left(j\frac{2\pi}{N_{\textrm{V}}}(\frac{N_{\textrm{V}}+1}{2}-n_{\textrm{V}})\eta(u)\right)\nonumber \\
 & \qquad=\frac{\sin\left(N_{\textrm{V}}(2\pi f_{m}h\cos(\theta_{\textrm{R},l})/c-2\pi\eta(u)/N_{\textrm{V}})/2\right)}{\sin\left((2\pi f_{m}h\cos(\theta_{\textrm{R},l})/c-2\pi\eta(u)/N_{\textrm{V}})/2\right)}\label{eq:AVm}
\end{align}
and
\begin{align}
[\tilde{\mathbf{a}}_{\textrm{H},m}(\phi_{\mathrm{\textrm{R}},l},\theta_{\mathrm{\textrm{R}},l})]_{p+P+1,1}=\frac{1}{\sqrt{N_{\textrm{H}}}}\mathbf{U}_{\textrm{sH}}^{H}\mathbf{a}_{\textrm{H},m}(\phi_{\mathrm{\textrm{R}},l},\theta_{\mathrm{\textrm{R}},l})\nonumber \\
=\frac{1}{\sqrt{N_{\textrm{H}}}}A_{\textrm{PM},p}\approx j^{p}J_{p}(\varpi_{m,l})e^{-jp\phi_{\textrm{R},l}}.\label{eq:AHm}
\end{align}
Steps 1 and 2 are indispensable, reducing the number of required RF
chains substantially from $N_{\textrm{R}}$ to $(2P+1)N_{\textrm{B}}$.

\subsection{Multidimensional Spatial Interpolation (MDSI)\label{subsec:Multidimensional-Spatial-Interpo}}

When the fractional bandwidth or the scale of the antenna array is
large, the aforementioned beam squint effect arises \cite{wideband}.
This is because the array response vectors \eqref{eq:AVm} and \eqref{eq:AHm}
depend on the frequency of the specific subcarrier $f_{m}$. The beam
squint effect would compromise the capability of jointly utilizing
the received signals at all frequency bands to estimate the path parameters.
As a result, the high temporal resolution of wideband mmWave systems
could not be effectively exploited.

One could keep the array response matrices consistent across all frequencies,
by transforming the array response vectors \eqref{eq:AVm} and \eqref{eq:AHm}
associated with the frequency $f_{m}$, $\forall m=0,1,\ldots,M-1$,
into the corresponding array response vectors at the reference frequency
$f_{0}$ \cite{Inter1}. For continuous signals, this could be ideally
achieved by the Shannon-Whittaker interpolation \cite{shannon}, which
sets different vertical distances and radii at different frequencies,
i.e., $h_{\textrm{vi},m}=f_{0}h/f_{m}$ and $r_{\textrm{vi},m}=f_{0}r/f_{m}$.
Then, from \eqref{eq:AVm} and \eqref{eq:AHm}, the virtual vertical
and horizontal response vectors, $\mathbf{\dot{a}}_{\textrm{V},m}(\theta_{\mathrm{\textrm{R}},l})$
and $\mathbf{\dot{a}}_{\textrm{H},m}(\theta_{\mathrm{\textrm{R}},l})$,
can be constructed as
\begin{align}
 & [\mathbf{\dot{a}}_{\textrm{V},m}(\theta_{\mathrm{\textrm{R}},l})]_{u,1}\nonumber \\
 & =\frac{\sin\left(N_{\textrm{V}}(2\pi f_{m}h_{\textrm{vi},m}\cos(\theta_{\textrm{R},l})/c-2\pi\eta(u)/N_{\textrm{V}})/2\right)}{\sin\left((2\pi f_{m}h_{\textrm{vi},m}\cos(\theta_{\textrm{R},l})/c-2\pi\eta(u)/N_{\textrm{V}})/2\right)}\nonumber \\
 & =\frac{\sin\left(N_{\textrm{V}}(2\pi f_{0}h\cos(\theta_{\textrm{R},l})/c-2\pi\eta(u)/N_{\textrm{V}})/2\right)}{\sin\left((2\pi f_{0}h\cos(\theta_{\textrm{R},l})/c-2\pi\eta(u)/N_{\textrm{V}})/2\right)}\nonumber \\
 & =[\tilde{\mathbf{a}}_{\textrm{V},0}(\theta_{\mathrm{\textrm{R}},l})]_{u,1}\label{eq:AVm0}
\end{align}
and
\begin{align}
 & [\mathbf{\dot{a}}_{\textrm{H},m}(\phi_{\mathrm{\textrm{R}},l},\theta_{\mathrm{\textrm{R}},l})]_{p+P+1,1}\nonumber \\
 & =j^{p}J_{p}(\frac{2\pi}{c}f_{m}r_{\textrm{vi},m}\sin(\theta_{\textrm{R},l}))e^{-jp\phi_{\textrm{R},l}}\nonumber \\
 & =j^{p}J_{p}(\frac{2\pi}{c}f_{0}r\sin(\theta_{\textrm{R},l}))e^{-jp\phi_{\textrm{R},l}}\nonumber \\
 &=[\tilde{\mathbf{a}}_{\textrm{H},0}(\phi_{\mathrm{\textrm{R}},l},\theta_{\mathrm{\textrm{R}},l})]_{p+P+1,1}.\label{eq:AHm0}
\end{align}
The signal reconstructed by using \eqref{eq:AVm0} and \eqref{eq:AHm0}
can be expressed as
\begin{align}
 & \mathbf{\dot{y}}_{\textrm{s2,}m}=\sum_{l=1}^{N_{\textrm{p}}}\beta_{l}x_{m}e^{-j2\pi f_{m}\tau_{l}}(\mathbf{\dot{a}}_{\textrm{V},m}(\theta_{\mathrm{\textrm{R}},l})\nonumber \\
 & \qquad\otimes\mathbf{\dot{a}}_{\textrm{H},m}(\phi_{\mathrm{\textrm{R}},l},\theta_{\mathrm{\textrm{R}},l}))+\mathbf{W}_{\textrm{s2,}m}^{H}\mathbf{n}_{m}\nonumber \\
 & =\sum_{l=1}^{N_{\textrm{p}}}\beta_{l}x_{m}e^{-j2\pi f_{m}\tau_{l}}(\tilde{\mathbf{a}}_{\textrm{V},0}(\theta_{\mathrm{\textrm{R}},l})\otimes\tilde{\mathbf{a}}_{\textrm{H},0}(\phi_{\mathrm{\textrm{R}},l},\theta_{\mathrm{\textrm{R}},l}))\nonumber \\
 & \qquad+\mathbf{W}_{\textrm{s2,}m}^{H}\mathbf{n}_{m}\nonumber \\
 & =\sum_{l=1}^{N_{\textrm{p}}}\beta_{l}x_{m}e^{-j2\pi f_{m}\tau_{l}}\tilde{\mathbf{a}}_{0}(\phi_{\mathrm{\textrm{R}},l},\theta_{\mathrm{\textrm{R}},l})+\mathbf{W}_{\textrm{s2,}m}^{H}\mathbf{n}_{m}.\label{eq:idealsignal}
\end{align}
In practice, the Shannon-Whittaker interpolation could hardly achieve
perfect signal reconstruction for time-limited signals, and it also
has a high computational complexity \cite{shannon}.

In this paper, we extend linear interpolation \cite{Inter2} (which
is a low-complexity and effective method for data point construction)
to the multidimensional spatial interpolation. The multidimensional
array response matrices consistent across all frequencies can be constructed
by using the received time-limited signals. By applying the linear
interpolation in both the vertical and horizontal spatial domains,
we can reconstruct the signal in \eqref{eq:sp2channel} and obtain
an approximation of \eqref{eq:idealsignal}. The reconstructed signal
is calculated as
\begin{align}
\left[\mathbf{\tilde{\mathbf{y}}}_{\textrm{s2,}m}\right]_{n_{\textrm{DS},2},1} & =\left[\mathbf{y}_{\textrm{s2,}m}\right]_{n_{\textrm{DS},2},1}+\frac{r_{\textrm{vi},m}}{r}\Delta_{\mathbf{y}_{\textrm{s2,H,}m}}+\frac{h_{\textrm{vi},m}}{h}\Delta_{\mathbf{y}_{\textrm{s2,V,}m}},\label{eq:multiDsignal}
\end{align}
where $n_{\textrm{DS},2}=(2P+1)(u-1)+p$. If $n_{\textrm{DS},2}\leq(2P+1)(N_{\textrm{B}}-1)$,
$\Delta_{\mathbf{y}_{\textrm{s2,H,}m}}$ and $\Delta_{\mathbf{y}_{\textrm{s2,V,}m}}$
are constructed as $\Delta_{\mathbf{y}_{\textrm{s2,H,}m}}=\left[\mathbf{y}_{\textrm{s2,}m}\right]_{(n_{\textrm{DS},2}+1),1}-\left[\mathbf{y}_{\textrm{s2,}m}\right]_{n_{\textrm{S}},1}$
and $\Delta_{\mathbf{y}_{\textrm{s2,V,}m}}=\left[\mathbf{y}_{\textrm{s2,}m}\right]_{(n_{\textrm{DS},2}+2P+1),1}-\left[\mathbf{y}_{\textrm{s2,}m}\right]_{n_{\textrm{DS},2},1}$,
respectively. Otherwise, $\Delta_{\mathbf{y}_{\textrm{s2,H,}m}}=\left[\mathbf{y}_{\textrm{s2,}m}\right]_{n_{\textrm{DS},2},1}-\left[\mathbf{y}_{\textrm{s2,}m}\right]_{(n_{\textrm{DS},2}-1),1}$
and $\Delta_{\mathbf{y}_{\textrm{s2,V,}m}}=\left[\mathbf{y}_{\textrm{s2,}m}\right]_{n_{\textrm{DS},2},1}-\left[\mathbf{y}_{\textrm{s2,}m}\right]_{(n_{\textrm{DS},2}-2P-1),1}.$

\section{Proposed Wideband Channel Parameter Estimation \label{sec:Proposed-Wideband-3D}}

In this section, we estimate the path parameters and the 3D position
of the MS based on the processed signals in Sections \ref{sec:Proposed-Two-Step-Wideband}
and \ref{subsec:Multidimensional-Spatial-Interpo}. Since the beamformers
developed in Section \ref{sec:Proposed-Two-Step-Wideband}  are linear
transforms, the multiple-invariance structure required for ESPRIT
can be recovered losslessly between respective submatrices of the
space-time response matrix. By exploiting the recurrence relations
in the multiple-invariance structure, the delay and elevation angle
of each path can be estimated using ESPRIT. For the azimuth angles,
the expression for the horizontal array response vectors \eqref{eq:AHm}
does not exhibit any recurrence. Hence the azimuth angles are estimated
by using MUSIC after obtaining the corresponding elevation angles.
The hardware and software complexities of the proposed joint delay
and angle  estimation approach are analyzed in the end.

\subsection{Wideband JADE Algorithm\label{subsec:Wideband-JADE-Algorithm}}

Collecting the received signals at all frequencies, we have $\mathbf{\tilde{\mathbf{y}}}=\left[\mathbf{\tilde{\mathbf{y}}}_{\textrm{s2,}1},\mathbf{\tilde{\mathbf{y}}}_{\textrm{s2,}2},\ldots,\mathbf{\tilde{\mathbf{y}}}_{\textrm{s2,}M}\right].$
Assume that the same signals are transmitted at all subcarriers. We
can vectorize $\mathbf{\mathbf{\tilde{\mathbf{y}}}}$ as
\begin{equation}
\mathbf{\mathbf{\tilde{\mathbf{y}}}_{\textrm{vec}}}=\textrm{vec}(\mathbf{\mathbf{\tilde{\mathbf{y}}}})=\left[\mathbf{\Gamma}\diamond\tilde{\mathbf{A}}\right]\mathbf{d}+\textrm{vec}(\mathbf{\mathbf{\tilde{\mathbf{n}}}})=\mathbf{U}\mathbf{d}+\tilde{\mathbf{n}}_{\textrm{v}},
\end{equation}
where $\tilde{\mathbf{A}}=\left[\tilde{\mathbf{a}}_{0}(\phi_{\mathrm{\textrm{R}},1},\theta_{\mathrm{\textrm{R}},1}),\ldots,\tilde{\mathbf{a}}_{0}(\phi_{\mathrm{\textrm{R}},N_{\textrm{p}}},\theta_{\mathrm{\textrm{R}},N_{\textrm{p}}})\right],$
$\mathbf{\mathbf{\tilde{\mathbf{n}}}}=\mathbf{W}_{\textrm{s2}}^{H}\left[\mathbf{n}_{1},\ldots,\mathbf{n}_{M}\right],$
$\left[\mathbf{\Gamma}\right]_{m,l}=e^{-j2\pi f_{m}\tau_{l}}$, and
$\mathbf{d}=x\left[\beta_{1},\beta_{2},\ldots,\beta_{N_{\textrm{p}}}\right]^{T}$.
Here, $\mathbf{U}\in\mathbb{C}^{N_{\textrm{DS},2}M\times N_{\textrm{p}}}$,
also known as the space-time response matrix in \cite{SCIVanderveen},
collects the set of AOAs and path delays. The covariance matrix of
$\mathbf{\tilde{\mathbf{y}}}_{\textrm{vec}}$ can be calculated as
\begin{equation}
\mathbf{R}_{\mathbf{\tilde{\mathbf{y}}}_{\textrm{vec}}}=\mathbb{E}\left\{ \mathbf{\tilde{\mathbf{y}}}_{\textrm{vec}}\mathbf{\tilde{\mathbf{y}}}_{\textrm{vec}}^{H}\right\} =\mathbb{\mathbf{U}}\mathbf{\Lambda}_{\mathbf{d}}\mathbf{U}^{H}+\mathbf{\sigma}_{\textrm{n}}^{2}\mathbf{I}_{(N_{\textrm{DS},2}M)},\label{eq:Rh1}
\end{equation}
where $\mathbf{\Lambda}_{\mathbf{d}}=\mathbb{E}\left\{ \mathbf{dd}^{H}\right\} $
is a diagonal matrix. The eigenvalue-decomposition (EVD) of $\mathbf{R}_{\mathbf{\tilde{\mathbf{y}}}_{\textrm{vec}}}$
can be obtained by
\begin{align}
\mathbf{R}_{\mathbf{\tilde{\mathbf{y}}}_{\textrm{vec}}} & =\left[\mathbf{\mathbf{E}_{\textrm{s}}},\mathbf{E}_{\textrm{n}}\right]\left[\begin{array}{cc}
\mathbf{\Sigma}_{\mathbf{\textrm{s}}} & \mathbf{0}_{N_{\textrm{p}}\times(N_{\textrm{DS},2}M-N_{\textrm{p}})}\\
\mathbf{0}_{(N_{\textrm{DS},2}M-N_{\textrm{p}})\times N_{\textrm{p}}} & \mathbf{\sigma}_{\textrm{n}}^{2}\mathbf{I}_{N_{\textrm{DS},2}M-N_{\textrm{p}}}
\end{array}\right]\nonumber \\
 & \qquad\times\left[\mathbf{\mathbf{\mathbf{E}_{\textrm{s}}}},\mathbf{E}_{\textrm{n}}\right]^{H}=\mathbf{\mathbf{E}_{\textrm{s}}}\mathbf{\Sigma}_{\mathbf{\textrm{s}}}\mathbf{E}_{\textrm{s}}^{H}+\mathbf{\sigma}_{\textrm{n}}^{2}\mathbf{E}_{\textrm{n}}\mathbf{E}_{\textrm{n}}^{H},\label{eq:Rh3}
\end{align}
where $\mathbf{E}_{\textrm{s}}\in\mathbf{\mathbb{C}}^{N_{\textrm{DS},2}M\times N_{\textrm{p}}}$
and $\mathbf{\mathbf{E}_{\textrm{n}}\in\mathbb{C}}^{N_{\textrm{DS},2}M\times(N_{\textrm{DS},2}M-N_{\textrm{p}})}$
correspond to the signal subspace and noise subspace, respectively.
$\mathbf{\Sigma}_{\mathbf{\textrm{s}}}\in\mathbf{\mathbb{R}}^{N_{\textrm{p}}\times N_{\textrm{p}}}$
is a diagonal matrix whose elements are the $N_{\textrm{p}}$ largest
eigenvalues of $\mathbf{R}_{\bar{\mathbf{h}}}$. Based on $\mathbf{E}_{\textrm{n}}\mathbf{E}_{\textrm{n}}^{H}+\mathbf{\mathbf{E}_{\textrm{s}}}\mathbf{E}_{\textrm{s}}^{H}=\mathbf{I}_{N_{\textrm{DS},2}M}$,
\eqref{eq:Rh3} can be rewritten as
\begin{equation}
\mathbf{R}_{\mathbf{\mathbf{\mathbf{\tilde{\mathbf{y}}}_{\textrm{vec}}}}}=\mathbf{\mathbf{E}_{\textrm{s}}}(\mathbf{\mathbf{\Sigma}_{\mathbf{\textrm{s}}}-\mathbf{\mathbf{\sigma}_{\textrm{n}}^{2}\mathbf{I}}}_{N_{\textrm{p}}})\mathbf{E}_{\textrm{s}}^{H}+\mathbf{\mathbf{\sigma}_{\textrm{n}}^{2}}\mathbf{I}_{N_{\textrm{DS},2}M}.\label{eq:Rh2}
\end{equation}
By setting \eqref{eq:Rh1} and \eqref{eq:Rh2} equal, we obtain
\begin{equation}
\mathbf{E}_{\textrm{s}}=\mathbf{UT},\label{eq:Es}
\end{equation}
where $\mathbf{T}\in\mathbb{C}^{N_{\textrm{p}}\times N_{\textrm{p}}}$
is a full rank matrix.

As discussed below, $\mathbf{U}$ in \eqref{eq:Es} has a multiple-invariance
structure with a linear recurrence relationship. The relationship
allows the use of the ESPRIT method to estimate the delay and elevation
angle of each path.

\subsubsection{Delay Estimation\label{subsec:Delay-Estimation}}

Define the delay-selection matrix as $\mathbf{J}_{\textrm{D}}=\textrm{diag}\left(\mathbf{J}_{\textrm{D},1},\ldots,\mathbf{J}_{\textrm{D},M}\right)\in\mathbb{R}^{M\times N_{\textrm{DS},2}M},$
where $\mathbf{J}_{\textrm{D},m}=\mathbf{1}_{N_{\textrm{DS},2}}^{T}$.
We can obtain the delay-related submatrix $\mathbf{U}_{\textrm{D}}=\mathbf{J}_{\textrm{D}}\mathbf{U}\in\mathbb{C}^{M\times N_{\textrm{p}}}$.
By defining $\tilde{\mathbf{J}}_{\textrm{D},m}=[\mathbf{0}_{1\times(m-1)},1,$
$\mathbf{0}_{1\times(M-m)}]\in\mathbb{R}^{1\times M}$, the delay-related
submatrix associated with the frequency $f_{m}$ can be calculated
as $\mathbf{U}_{\textrm{D},m}=\tilde{\mathbf{J}}_{\textrm{D},m}\mathbf{U}_{\textrm{D}}\in\mathbb{C}^{1\times N_{\textrm{p}}}$.
Thus, we obtain a linear recurrence relation between the delay-related
submatrices of each frequency as
\begin{equation}
\mathbf{U}_{\textrm{D},\tilde{m}+1}=\mathbf{U}_{\textrm{D},\tilde{m}}\mathbf{\mathbf{\Theta}_{\textrm{D}}},\label{eq:ud}
\end{equation}
where $\mathbf{\mathbf{\Theta}_{\textrm{D}}}=\textrm{diag}\left(e^{-j2\pi\Delta_{\textrm{F}}\tau_{1}},\ldots,e^{-j2\pi\Delta_{\textrm{F}}\tau_{N_{\textrm{p}}}}\right)\in\mathbb{C}^{N_{\textrm{p}}\times N_{\textrm{p}}}$
and $\tilde{m}=1,2,\ldots,M-1.$

According to \eqref{eq:Es}, the delay-related submatrix of the signal
subspace matrix at the frequency $f_{m}$ can be given by
\begin{equation}
\mathbf{E}_{\textrm{D},m}=\tilde{\mathbf{J}}_{\textrm{D},m}\mathbf{J}_{\textrm{D}}\mathbf{\mathbf{E}_{\textrm{s}}}=\mathbf{U}_{\textrm{D},m}\mathbf{T}.\label{eq:42-1}
\end{equation}
Substituting \eqref{eq:ud} into \eqref{eq:42-1}, we obtain
\begin{equation}
\mathbf{E}_{\textrm{D},\tilde{m}+1}=\mathbf{E}_{\textrm{D},\tilde{m}}\mathbf{T}^{-1}\mathbf{\mathbf{\Theta}_{\textrm{D}}T}=\mathbf{E}_{\textrm{D},\tilde{m}}\mathbf{\Psi_{\textrm{D}}}.\label{eq:delay}
\end{equation}

By using the total least-squares (TLS) criterion \cite{38}, we estimate
$\Psi_{\textrm{D}}=\mathbf{T}^{-1}\mathbf{\mathbf{\Theta}_{\textrm{D}}T}=\mathbf{E}_{\textrm{D},\tilde{m}}^{\dagger}\mathbf{E}_{\textrm{D},\tilde{m}+1}$
as $\hat{\mathbf{\Psi}}_{\textrm{D},\tilde{m}}$, each of which has
a total of $N_{\textrm{p}}$ sorted eigenvalues, i.e., $\lambda_{\textrm{D},\tilde{m},N_{\textrm{p}}}$.
Due to the fact that the eigenvalues of an upper triangular matrix
are also diagonal elements of the matrix, we can obtain $(M-1)$ different
estimates for each $\mathbf{\mathbf{\Theta}_{\textrm{D}}}$. As a
result, the delay of the $n_{\textrm{p}}$-th path, $\tau_{n_{\textrm{p}}}$,
can be estimated as
\begin{equation}
\hat{\tau}_{l}=\frac{1}{M-1}\sum_{\tilde{m}}^{M-1}\left[j\ln(\lambda_{\textrm{D},\tilde{m},l})/2\pi\Delta_{\textrm{F}}\right].
\end{equation}

\subsubsection{Angle Estimation\label{subsec:Angle-Estimation}}

We first use the processed vertical array response vector \eqref{eq:AVm0}
to estimate the elevation angle. According to \eqref{eq:AVm}, the
$(u+1)$-th element of $\tilde{\mathbf{a}}_{\textrm{V},0}(\theta_{\mathrm{\textrm{R}},l})$
can be given by
\begin{align}
 & [\tilde{\mathbf{a}}_{\textrm{V},0}(\theta_{\mathrm{\textrm{R}},l})]_{u+1,1}\nonumber \\
 & =\frac{\sin\left(N_{\textrm{V}}(2\pi f_{0}h\cos(\theta_{\textrm{R},l})/c-2\pi\eta(u+1)/N_{\textrm{V}})/2\right)}{\sin\left((2\pi f_{0}h\cos(\theta_{\textrm{R},l})/c-2\pi\eta(u+1)/N_{\textrm{V}})/2\right)}.
\end{align}
Comparing $[\tilde{\mathbf{a}}_{\textrm{V},0}(\theta_{\mathrm{\textrm{R}},l})]_{u+1,1}$
with the $u$-th element in \eqref{eq:AVm0}, we see that two successive
components of the processed vertical array response vector, $[\tilde{\mathbf{a}}_{\textrm{V},0}(\theta_{\mathrm{\textrm{R}},l})]_{u,1}$
and $[\tilde{\mathbf{a}}_{\textrm{V},0}(\theta_{\mathrm{\textrm{R}},l})]_{u+1,1}$,
are related as follows.
\begin{align}
 & (-1)^{\eta(u)}\sin\left((g(\theta_{\textrm{R},l})-2\pi\eta(u)/N_{\textrm{V}})/2\right)[\tilde{\mathbf{a}}_{\textrm{V},0}(\theta_{\mathrm{\textrm{R}},l})]_{u,1}\nonumber \\
 & =(-1)^{\eta(u+1)}\sin\left((g(\theta_{\textrm{R},l})-2\pi\eta(u+1)/N_{\textrm{V}})/2\right)\nonumber \\
 & \qquad\times[\tilde{\mathbf{a}}_{\textrm{V},0}(\theta_{\mathrm{\textrm{R}},l})]_{u+1,1},\label{eq:AVuAVuq}
\end{align}
where $g(\theta_{\textrm{R},l})=\frac{2\pi}{c}f_{0}h\cos(\theta_{\textrm{R},l})$.
By trigonometric manipulations, we rewrite \eqref{eq:AVuAVuq} as
\begin{align}
 & (-1)^{\eta(u)}\sin\left(\eta(u)\frac{\pi}{N_{\textrm{V}}}\right)[\tilde{\mathbf{a}}_{\textrm{V},0}(\theta_{\mathrm{\textrm{R}},l})]_{u,1}\nonumber \\
 & \qquad+(-1)^{\eta(u+1)+1}\sin\left(\eta(u+1)\frac{\pi}{N_{\textrm{V}}}\right)[\tilde{\mathbf{a}}_{\textrm{V},0}(\theta_{\mathrm{\textrm{R}},l})]_{u+1,1}\nonumber \\
 & =\tan\left(\frac{g(\theta_{\textrm{R},l})}{2}\right)\left[(-1)^{\eta(u)}\cos\left(\eta(u)\frac{\pi}{N_{\textrm{V}}}\right)[\tilde{\mathbf{a}}_{\textrm{V},0}(\theta_{\mathrm{\textrm{R}},l})]_{u,1}\right.\nonumber \\
 & \qquad\left.+(-1)^{\eta(u+1)+1}\cos\left(\eta(u+1)\frac{\pi}{N_{\textrm{V}}}\right)[\tilde{\mathbf{a}}_{\textrm{V},0}(\theta_{\mathrm{\textrm{R}},l})]_{u+1,1}\right].
\end{align}
Stacking all $(N_{\textrm{B}}-1)$ equations together yields
\begin{equation}
\tan\left(\frac{g(\theta_{\textrm{R},l})}{2}\right)\mathbf{F}_{0}\tilde{\mathbf{a}}_{\textrm{V},0}(\theta_{\mathrm{\textrm{R}},l})=\mathbf{F}_{1}\tilde{\mathbf{a}}_{\textrm{V},0}(\theta_{\mathrm{\textrm{R}},l}),\label{eq:beamrelation}
\end{equation}
where
\begin{align*}
 & \left[\mathbf{F}_{0}\right]_{\tilde{u},u}=\\
 & \begin{cases}
(-1)^{\eta(\tilde{u})}\cos(2\pi\eta(\tilde{u})/N_{\textrm{V}}), & \textrm{if }u=\eta(\tilde{u});\\
(-1)^{\eta(\tilde{u}+1)+1}\cos(2\pi\eta(\tilde{u}+1)/N_{\textrm{V}}), & \textrm{if }u=\eta(\tilde{u}+1);\\
0, & \textrm{otherwise.}
\end{cases}
\end{align*}
\begin{align*}
 & \left[\mathbf{F}_{1}\right]_{\tilde{u},u}=\\
 & \begin{cases}
(-1)^{\eta(\tilde{u})}\sin(2\pi\eta(\tilde{u})/N_{\textrm{V}}), & \textrm{if }u=\eta(\tilde{u});\\
(-1)^{\eta(\tilde{u}+1)+1}\sin(2\pi\eta(\tilde{u}+1)/N_{\textrm{V}}), & \textrm{if }u=\eta(\tilde{u}+1);\\
0, & \textrm{otherwise.}
\end{cases}
\end{align*}
with $\tilde{u}=1,2,\ldots,N_{\textrm{B}}-1$.

The processes of selecting the angle-related submatrices are similar
to that of selecting the delay-related submatrices. Define the angle
selection matrix as $\mathbf{J}_{\textrm{A}}=\mathbf{1}_{M}^{T}\otimes\mathbf{I}_{N_{\textrm{DS},2}}\in\mathbb{R}^{N_{\textrm{DS},2}\times N_{\textrm{DS},2}M}$.
Then the angle-related submatrix can be formulated as $\mathbf{U}_{\textrm{A}}=\mathbf{J}_{\textrm{A}}\mathbf{U}\in\mathbb{C}^{N_{\textrm{DS},2}\times N_{\textrm{p}}}$.
Based on the recurrence relation in \eqref{eq:beamrelation}, we can
construct
\begin{equation}
\mathbf{F}_{0}\mathbf{U}_{\textrm{V}}\mathbf{\mathbf{\Theta}_{\textrm{V}}}=\mathbf{F}_{1}\mathbf{U}_{\textrm{V}},\label{eq:ANGLERELATION}
\end{equation}
where $\mathbf{\mathbf{\Theta}_{\textrm{V}}}=\textrm{diag}\left(\tan(g(\theta_{\textrm{R},1})/2),\ldots,\tan(g(\theta_{\textrm{R},N_{\textrm{p}}})/2)\right)$
and $\mathbf{U}_{\textrm{V}}=\mathbf{J}_{\textrm{V}}\mathbf{U}_{\textrm{A}}\in\mathbb{C}^{N_{\textrm{B}}\times N_{\textrm{p}}}$
is a submatrix of $\mathbf{U}_{\textrm{A}}$, where $\mathbf{J}_{\textrm{V}}=\mathbf{I}_{(2P+1)}\otimes\mathbf{1}_{N_{\textrm{B}}}^{T}\in\mathbb{R}^{N_{\textrm{B}}\times N_{\textrm{DS},2}}$.
Thus, the vertical array response-related submatrix can be calculated
as
\begin{equation}
\mathbf{E}_{\textrm{V}}=\mathbf{J}_{\textrm{V}}\mathbf{J}_{\textrm{A}}\mathbf{\mathbf{E}_{\textrm{s}}}=\mathbf{J}_{\textrm{V}}\mathbf{J}_{\textrm{A}}\mathbf{UT}=\mathbf{U}_{\textrm{V}}\mathbf{T}.\label{eq:anglesub}
\end{equation}
Substituting \eqref{eq:anglesub} into \eqref{eq:ANGLERELATION},
we can obtain
\begin{equation}
\mathbf{F}_{0}\mathbf{E}_{\textrm{V}}\mathbf{T}^{-1}\mathbf{\mathbf{\Theta}_{\textrm{V}}T}=\mathbf{F}_{0}\mathbf{E}_{\textrm{V}}\mathbf{\Psi_{\textrm{V}}}=\mathbf{F}_{1}\mathbf{E}_{\textrm{V}}.\label{eq:elementangle}
\end{equation}

With reference to the delay estimation in Section \ref{subsec:Delay-Estimation},
the elevation angle of the $l$-th path, $\hat{\theta}_{l}$, can
be estimated as
\begin{equation}
\hat{\theta}_{\mathrm{\textrm{R}},l}=\arccos\left(\arctan(\lambda_{\textrm{V},l})/\pi f_{0}h\right),
\end{equation}
where $\lambda_{\textrm{V},l}$ is the $l$-th eigenvalue of $\hat{\mathbf{\Psi}}_{\textrm{V}}$,
and $\hat{\mathbf{\Psi}}_{\textrm{V}}$ is the estimated matrix of
$\mathbf{\Psi_{\textrm{V}}}=\mathbf{T}^{-1}\mathbf{\mathbf{\Theta}_{\textrm{V}}T}$.

According to \eqref{eq:AHm}, the expression for each horizontal response
vector, which does not have the invariance structure, is an exponential
function weighted by the Bessel function. There is no recursive relationship
for the azimuth angle estimation. After obtaining the elevation angles,
we use MUSIC to estimate their corresponding azimuth angles.

Define $\mathbf{J}_{\textrm{H}}=\mathbf{J}_{\textrm{HA}}\mathbf{J}_{\textrm{A}}\in\mathbb{R}^{(2P+1)\times N_{\textrm{DS},2}M}$,
where $\mathbf{J}_{\textrm{HA},u}=\left[\mathbf{I}_{(2P+1)},\mathbf{0}_{(2P+1)\times(2P+1)(N_{\textrm{B}}-1)}\right]\in\mathbb{R}^{(2P+1)\times N_{\textrm{DS},2}}.$
We can obtain the corresponding horizontal signal $\mathbf{\mathbf{\tilde{\mathbf{y}}}_{\textrm{vec,H}}}=\mathbf{J}_{\textrm{H}}\mathbf{\mathbf{\tilde{\mathbf{y}}}_{\textrm{vec}}}\in\mathbb{C}^{(2P+1)\times1}$.
As done in \eqref{eq:Rh3}, the covariance matrix of $\mathbf{\mathbf{\mathbf{\tilde{\mathbf{y}}}_{\textrm{vec,H}}}}$
can be calculated as $\mathbf{R}_{\mathbf{\mathbf{\tilde{\mathbf{y}}}_{\textrm{vec,H}}}}=\mathbf{\mathbf{E}_{\textrm{s,H}}}\mathbf{\Sigma}_{\mathbf{\textrm{s,H}}}\mathbf{E}_{\textrm{s,H}}^{H}+\mathbf{\sigma}_{\textrm{n}}^{2}\mathbf{E}_{\textrm{n,H}}\mathbf{E}_{\textrm{n,H}}^{H},$
where $\mathbf{E}_{\textrm{s,H}}$ and $\mathbf{E}_{\textrm{n,H}}$
are the signal and noise subspaces of $\mathbf{\mathbf{\tilde{\mathbf{y}}}_{\textrm{vec,H}}}$,
respectively.

By substituting the estimate of the $l$-th path, $\hat{\theta}_{\mathrm{\textrm{R}},l}$,
in the MUSIC estimator, the azimuth angle of the path can be estimated
by
\begin{equation}
\hat{\phi}_{\mathrm{\textrm{R}},l}=\arg\max_{\varPhi}\left\Vert \mathbf{E}_{\textrm{n,H}}^{H}\tilde{\mathbf{a}}_{\textrm{H},0}(\varPhi,\hat{\theta}_{\mathrm{\textrm{R}},l})\right\Vert _{\textrm{F}}^{-2},\label{eq:fi}
\end{equation}
where $\varPhi$ is the azimuth of the AOA, and can be estimated by
1D search.

\subsection{Multipath Parameter Matching\label{subsec:Multipath-Parameter-Matching}}

As described above, the estimated channel parameters of each path
can be matched automatically in the absence of noises. This is because
they have the common factor $\mathbf{T}$, as shown in \eqref{eq:Es}.
In the presence of non-negligible noises, there can be a mismatch
between the estimated parameters. We take the delay and the elevation
AOA for an example. According to \eqref{eq:delay} and \eqref{eq:elementangle},
we have $\mathbf{\Psi}_{\textrm{D}}=\mathbf{T}_{\textrm{D}}^{-1}\mathbf{\mathbf{\Theta}_{\textrm{D}}}\mathbf{T}_{\textrm{D}}$
and $\mathbf{\Psi_{\textrm{V}}}=\mathbf{T}_{\textrm{V}}^{-1}\mathbf{\mathbf{\Theta}_{\textrm{V}}}\mathbf{T}_{\textrm{V}}$,
but $\mathbf{T}_{\textrm{V}}\neq\mathbf{T}_{\textrm{D}}\neq\mathbf{T}$
because of the noise. Most existing pair matching methods would require
the approximate values of the estimates first, and then use exhaustive
search to match all possible parameter pairs \cite{we2}. Such methods
would incur a prohibitive computational complexity if the numbers
of paths and parameters are large.

We note that in our approach, the estimated elevation angles, $\hat{\theta}_{\mathrm{\textrm{R}},l}$,
are used for the estimation of the azimuth angles, $\hat{\phi}_{\mathrm{\textrm{R}},l}$,
so that the azimuth and elevation angles of each path always match;
see \eqref{eq:fi}. However, there is a mismatch between the estimated
delays and angles, primarily caused by the noises. The
mismatch between the estimated delays and angles can be mitigated
by suppressing misalignment between the eigenvalues of the two matrices
$\mathbf{\Psi}_{\textrm{D}}$ and $\mathbf{\Psi}_{\textrm{V}}$. To
achieve this, we introduce two perturbation terms, $\mathbf{P}_{\textrm{D}}$
and $\mathbf{P}_{\textrm{V}}$, to address the potential misalignment
(resulting from the non-negligible receive noises) between the eigenvalues
of the two matrices $\mathbf{\Psi}_{\textrm{D}}$ and $\mathbf{\Psi}_{\textrm{V}}$,
hence pairing the estimated delays and angles for every path. $\mathbf{P}_{\textrm{D}}$
denotes the difference between the estimated delay eigenvalue matrix
in the presence of the noises, $\mathbf{\Psi}_{\textrm{D}}$, and
the actual delay eigenvalue matrix in the absence of the noises, $\tilde{\mathbf{\Psi}}_{\textrm{D}}$.
Therefore, $\tilde{\mathbf{\Psi}}_{\textrm{D}}=\mathbf{\Psi}_{\textrm{D}}+\mathbf{P}_{\textrm{D}}$.
Likewise, $\mathbf{P}_{\textrm{V}}$ denotes the difference between
the estimated angle eigenvalue matrix in the presence of the noises,
$\mathbf{\Psi}_{\textrm{V}}$, and its noise-free counterpart, $\tilde{\mathbf{\Psi}}_{\textrm{V}}$.
Therefore, $\tilde{\mathbf{\Psi}}_{\textrm{V}}=\mathbf{\Psi}_{\textrm{V}}+\mathbf{P}_{\textrm{V}}$.
From \eqref{eq:delay} and \eqref{eq:elementangle}, $\tilde{\mathbf{\Psi}}_{\textrm{D}}=\mathbf{\Psi}_{\textrm{D}}+\mathbf{P}_{\textrm{D}}=\mathbf{\tilde{T}}_{\textrm{D}}^{-1}\mathbf{\mathbf{\Theta}_{\textrm{D}}}\tilde{\mathbf{T}}_{\textrm{D}}$
and $\tilde{\mathbf{\Psi}}_{\textrm{V}}=\mathbf{\Psi}_{\textrm{V}}+\mathbf{P}_{\textrm{V}}=\mathbf{\tilde{T}}_{\textrm{V}}^{-1}\mathbf{\mathbf{\Theta}_{\textrm{V}}}\tilde{\mathbf{T}}_{\textrm{V}}$.
As discussed at the beginning of this subsection, the eigenvalues
of $\tilde{\mathbf{\Psi}}_{\textrm{D}}$ and $\tilde{\mathbf{\Psi}}_{\textrm{V}}$
match perfectly under the ideal, noise-free situation. Therefore,
by evaluating $\mathbf{P}_{\textrm{D}}$ and $\mathbf{P}_{\textrm{V}}$
to minimize the mismatch between the eigenvalues of $\mathbf{\Psi}_{\textrm{D}}+\mathbf{P}_{\textrm{D}}$
and $\mathbf{\Psi}_{\textrm{V}}+\mathbf{P}_{\textrm{V}}$, we can
obtain $\tilde{\mathbf{T}}_{\textrm{D}}=\tilde{\mathbf{T}}_{\textrm{V}}=\tilde{\mathbf{T}}$.
The estimated delays and angles can be correctly paired for different
paths; in other words, the parameter pair matching in \eqref{eq:delay}
and \eqref{eq:elementangle} can be achieved. The perturbation matrices
$\mathbf{P}_{\textrm{D}}$ and $\mathbf{P}_{\textrm{V}}$ can be obtained
by solving the following problem \cite{pair}:
\begin{align}
\min_{P_{\textrm{D}},P_{\textrm{V}}} & \left\Vert \mathbf{P}_{\textrm{D}}\right\Vert _{\mathrm{F}}^{2}+\left\Vert \mathbf{P}_{\textrm{V}}\right\Vert _{\mathrm{F}}^{2}\label{eq:matching}\\
s.t. & \left(\mathbf{\Psi}_{\textrm{D}}+\mathbf{P}_{\textrm{D}}\right)\left(\mathbf{\Psi}_{\textrm{V}}+\mathbf{P}_{\textrm{V}}\right)=\left(\mathbf{\Psi}_{\textrm{D}}+\mathbf{P}_{\textrm{V}}\right)\left(\mathbf{\Psi}_{\textrm{V}}+\mathbf{P}_{\textrm{D}}\right),\label{eq:mat2}
\end{align}
where \eqref{eq:matching} is formulated due to the fact that $\mathbf{P}_{\textrm{D}}$
and $\mathbf{P}_{\textrm{V}}$ need to obey the minimum Frobenius
norm constraint \cite{pair}. The exact solution to this non-linearly
constrained problem \eqref{eq:matching} is hard to find. To solve
the problem, we rewrite \eqref{eq:mat2} as
\begin{align}
 & \mathbf{P}_{\textrm{D}}\mathbf{P}_{\textrm{V}}-\mathbf{P}_{\textrm{V}}\mathbf{P}_{\textrm{D}}\nonumber \\
 & =\mathbf{\Psi}_{\textrm{D}}\mathbf{\Psi}_{\textrm{V}}+\mathbf{P}_{\textrm{V}}\mathbf{\Psi}_{\textrm{V}}+\mathbf{\Psi}_{\textrm{D}}\mathbf{P}_{\textrm{D}}-\mathbf{\Psi}_{\textrm{D}}\mathbf{\Psi}_{\textrm{V}}-\mathbf{\Psi}_{\textrm{D}}\mathbf{P}_{\textrm{V}}-\mathbf{P}_{\textrm{D}}\mathbf{\Psi}_{\textrm{V}}.\label{eq:mat3}
\end{align}
We assume that the perturbations are much smaller than $\mathbf{\Psi}_{\textrm{D}}$
and $\mathbf{\Psi}_{\textrm{V}}$, then the term $(\mathbf{P}_{\textrm{D}}\mathbf{P}_{\textrm{V}}-\mathbf{P}_{\textrm{V}}\mathbf{P}_{\textrm{D}})$
in \eqref{eq:mat3} can be suppressed \cite{mat}.

We can fix one of the two eigenvalue matrices (e.g.,
$\mathbf{P}_{\textrm{V}}$) and match the other (e.g., $\mathbf{P}_{\textrm{D}}$)
against it. To this end, we set $\mathbf{P}_{\textrm{V}}=\mathbf{0}$
and focus our evaluation on $\mathbf{P}_{\textrm{D}}$. By setting
$\mathbf{P}_{\textrm{V}}=\mathbf{0}$, $\mathbf{P}_{\textrm{D}}$
can be obtained as
\begin{align}
\textrm{vec}(\mathbf{P}_{\textrm{D}})=[\mathbf{\mathbf{\Psi}}_{\textrm{V}}^{T}\oplus(-\mathbf{\Psi}_{\textrm{V}})]^{\dagger}\textrm{vec}(\mathbf{\Psi}_{\textrm{V}}\mathbf{\Psi}_{\textrm{D}}-\mathbf{\Psi}_{\textrm{D}}\mathbf{\Psi}_{\textrm{V}}).\label{eq:pair}
\end{align}
By adding the perturbation matrix $\mathbf{P}_{\textrm{D}}$ to the
elevation angle eigenvalue matrix $\mathbf{\Psi}_{\textrm{D}}$, the
delay and the elevation angles can be matched. The parameters of each
path can be associated correctly. It is worth pointing
out that it is dramatically simpler to only evaluate $\mathbf{P}_{\textrm{D}}$
in problem \eqref{eq:pair} than it is to evaluate both $\mathbf{P}_{\textrm{D}}$
and $\mathbf{P}_{\textrm{V}}$ in problem \eqref{eq:matching}. This
is because \eqref{eq:matching} is a non-linearly constrained problem.

\subsection{3D Localization Based on Estimated Channel Parameters\label{subsec:Positioning-Method}}

Given the estimates of the azimuth and elevation
AOAs, and the propagation delay of every path, we can specify the
3D direction and the (relative) length of the path. With the knowledge
of the physical environments\footnote{This knowledge can be acquired by using existing techniques, such
as coded structured light-based 3D reconstruction \cite{3Dmeasurement}
and multi-viewpoint cloud matching \cite{3Dmeasurement2}. The parameters
of reflection/refraction paths from static objects can also be extracted
from long-term estimates.}, the MS can be accordingly located. In the case
that the time offset between the BS and the MS, $\tau_{\textrm{of}}$,
is known to the BS, a single line-of-sight (LOS) or non-LOS (NLOS)
path suffices to locate the MS by retrospectively tracing along the
estimated direction of the path (starting from the BS) for the estimated
signal propagation distance. In the case that $\tau_{\textrm{of}}$
is unknown, the distance over which an impinging signal propagates
along a path before reaching the BS may not be accurate. At least
two paths are needed. In the ideal (noise-free) scenario, the two
paths intersect twice. One of the intersections is the BS, and the
other indicates the location of the MS. In the more practical scenario
with non-negligible noises, the two estimated paths may not intersect,
except at the BS. The MS can be estimated to be at such a position
that: its projections on the two paths account for the least squared
difference from the estimated delay difference of the paths while
the total of its squared distances to the projections is the minimum
(e.g., by minimizing the (weighted) sum of the squared difference
and distances).

\subsection{Complexity Analysis\label{subsec:Complexity-Analysis}}

\begin{figure}
\centering{}\centering \includegraphics[width=7cm]{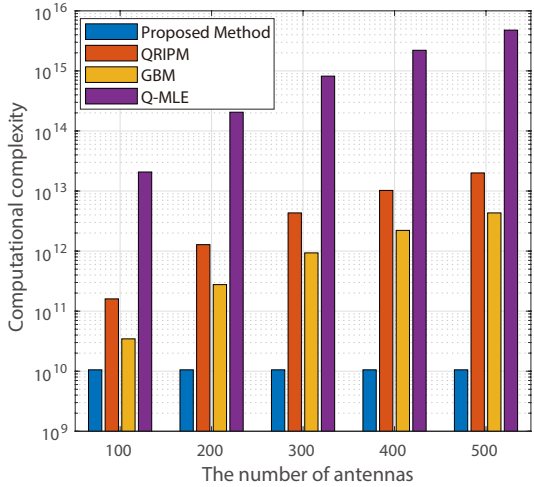} \caption{Variation of the computational complexity vs. the number of antennas.\label{fig:Complex}}
\end{figure}

We proceed to analyze the hardware and software complexities of the
proposed joint delay and angle  estimation approach. For a large-scale
antenna array system using fully digital beamforming, its hardware
complexity is $O(N_{\textrm{R}})$. In our proposed approach, the
use of the hybrid beamformer allows for a dramatic reduction of the
hardware complexity from $O(N_{\textrm{R}})$ to $O(N_{\textrm{RF}})$,
where $N_{\textrm{RF}}=\max(N_{\textrm{V}},(2P+1)N_{\textrm{B}})$.

In terms of signal processing complexity, we compare the proposed
approach with the state-of-the-art techniques, namely, GBM \cite{we11},
QRIPM \cite{34}, and Q-MLE \cite{QMLE}. For the proposed approach,
after hybrid beamforming, the dimension of the received signal is
reduced to $N_{\textrm{DS},2}$, so the computational complexity of
MDSL processing is $O(N_{\textrm{DS},2}M)=O((2P+1)N_{\textrm{B}}M)=O(\gamma PN_{\textrm{p}}M)$.
The computational complexities of calculating the covariance matrix,
$\mathbf{R}_{\mathbf{\tilde{\mathbf{y}}}_{\textrm{vec}}}$, in \eqref{eq:Rh1}
and performing the EVD on $\mathbf{R}_{\mathbf{\tilde{\mathbf{y}}}_{\textrm{vec}}}$
according to \eqref{eq:Rh3} are $O(\gamma^{2}P^{2}N_{\textrm{p}}^{2}M^{2}T_{s})$
and $O(\gamma^{3}P^{3}N_{\textrm{p}}^{3}M^{3})$, respectively, where
$T_{s}$ is the number of snapshots. The complexities of computing
the delay $\hat{\tau}_{l}$ and the elevation angle, $\hat{\theta}_{\mathrm{\textrm{R}},l}$,
are $O(MN_{\textrm{p}}^{3})$ and $O(\gamma^{2}N_{\textrm{p}}^{2}+N_{\textrm{p}}^{3})$,
respectively. When estimating $\hat{\phi}_{\mathrm{\textrm{R}},l}$
with 1D search using \eqref{eq:fi}, the computational complexity
is $O(\gamma^{2}N_{\textrm{p}}^{2}D)$, where $D$ is the size of
the search dimension. For the pair matching operation, the computational
complexity is $O(N_{\textrm{p}}^{3})$. Thus, the overall computational
complexity of our proposed approach is $O(\gamma PN_{\textrm{p}}M+\gamma^{2}P^{2}N_{\textrm{p}}^{2}M^{2}T_{s}+\gamma^{3}P^{3}N_{\textrm{p}}^{3}M^{3}+MN_{\textrm{p}}^{3}+\gamma^{2}N_{\textrm{p}}^{2}+N_{\textrm{p}}^{3}+\gamma^{2}N_{\textrm{p}}^{2}D+N_{\textrm{p}}^{3})$,
which does not depend on the number of receive antennas $N_{\textrm{R}}$.
The computational complexities of QRIPM and GBM increase rapidly,
as the number of receive antennas increases. When the number of receive
antennas $N_{\textrm{R}}$ is large, the computational complexities
of QRIPM and GBM are $O(N_{\textrm{R}}^{3}M^{4})$ and $O(P^{3}N_{\textrm{V}}^{3}M^{4})$,
respectively. The computational complexity of Q-MLE is $O(N_{\textrm{R}}^{2}M^{2}N_{\textrm{AZI}}N_{\textrm{ELE}}N_{\textrm{DEL}}+(N_{\textrm{p}}N_{\textrm{R}}M)^{3.5})$,
where $N_{\textrm{AZI}}$, $N_{\textrm{ELE}}$, and $N_{\textrm{DEL}}$
are the search grids of azimuth angle, elevation angle, and delay,
respectively.

Fig. \ref{fig:Complex} compares the computational complexities of
the four methods with the growing number of antennas $N_{\textrm{R}}=N_{\textrm{H}}N_{\textrm{V}}$,
where $\gamma=2$, $N_{\textrm{p}}=3$, $M=20$, and $P=12$. We set
$D=N_{\textrm{AZI}}=N_{\textrm{ELE}}=N_{\textrm{DEL}}=100$. The figure
shows that, compared with the existing methods, the proposed approach
has a substantially lower computational complexity. The gaps between
the proposed algorithm and the existing alternatives are increasingly
significant with the growing number of receive antennas at the BS.

\section{Simulation Results\label{sec:Simulation-Results}}

\begin{figure*}
\begin{centering}
\includegraphics[width=16cm]{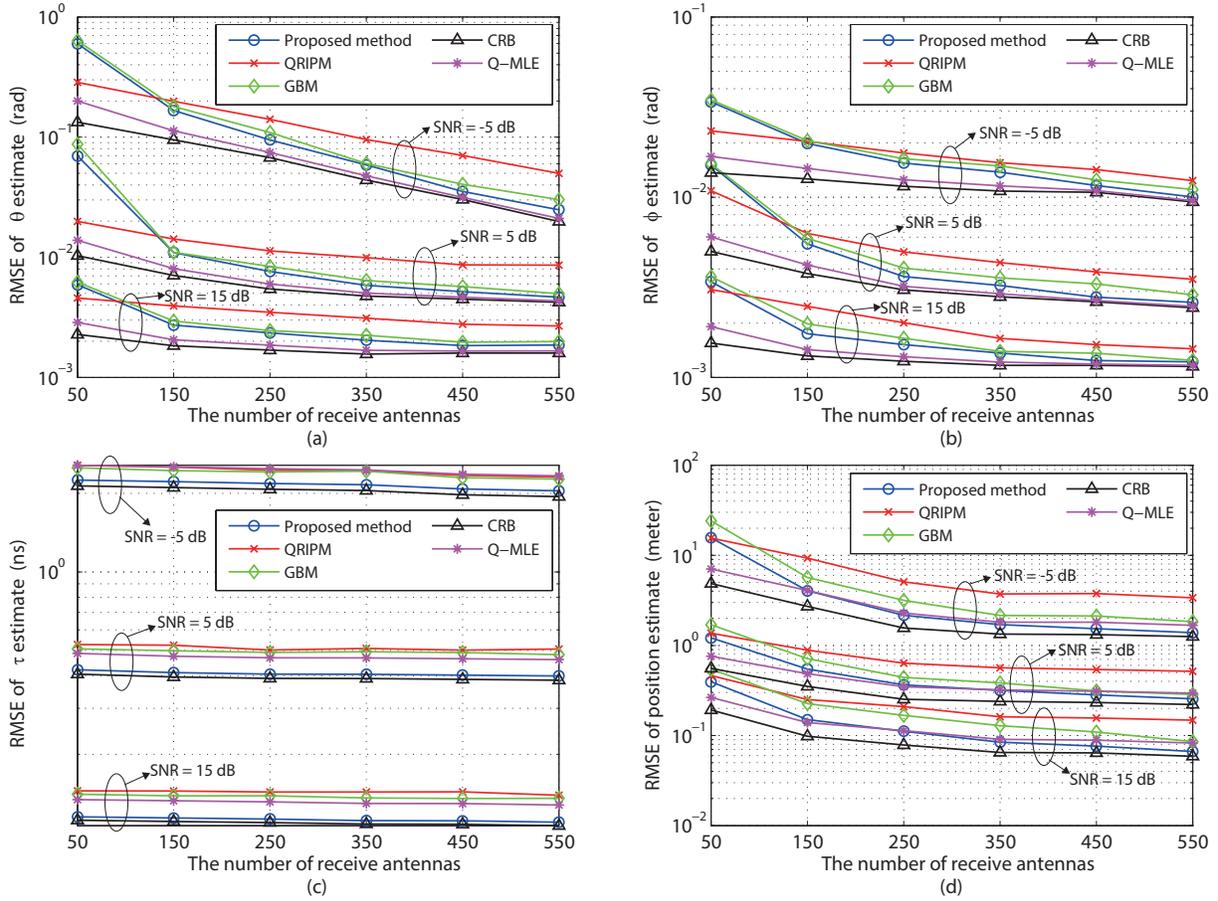}
\par\end{centering}
\centering{}\caption{The RMSE vs. the number of BS antennas for the estimation of different
parameters. (a) Azimuth AOA; (b) Elevation AOA; (c) Path delay; and
(d) MS position.\label{fig:result0}}
\end{figure*}

In this section, we present simulation results to demonstrate the
performance of the proposed approach under different parameters. We
set $f_{0}=30$ GHz and $B=2$ GHz\footnote{The beam squint depends on both the fractional bandwidth and the scale
of the deployed antenna array \cite{wideband}. In the case of 30
GHz carrier frequency and 2 GHz bandwidth, the fractional bandwidth
is 0.067, which is non-negligible and can result in a noticeable beam
squint, especially when the size of the array is large. }, and assume that there are a total of $N_{\textrm{p}}=3$ NLOS paths
and $M=20$ consecutive subcarriers. The distance, $h$, between adjacent
receiving UCAs and the radius, $r$, of each UCA are $0.5\lambda_{0}$
and $2\lambda_{0}$, respectively.

Fig. \ref{fig:result0} plots the root mean square
errors (RMSEs) of the estimated angle, delay, and MS position with
the increasing number of receive antennas, under different SNR conditions.
The proposed algorithm is compared with GBM \cite{we11}, QRIPM \cite{34},
Q-MLE \cite{QMLE}, and the  Cram\'{e}r-Rao lower bound (CRLB)\footnote{The CRLB is calculated according to \cite{bounds}.}.
Note that GBM \cite{we11}, QRIPM \cite{34}, and Q-MLE \cite{QMLE}
are the state of the art for solving the considered parameter estimation
problem for UCyAs, and act as the benchmarks in this paper. In particular,
the sparsity of the channel is exploited in \cite{we11}, where the
angular space is first discretized and then discrete directions with
significant incoming powers are picked up to estimate the channel
parameters.

Despite sparse representation techniques were also
developed to exploit the sparsity of mmWave multi-antenna systems
for channel estimation in \cite{Shahmansoori} and \cite{Re2LilongDai},
the techniques are not applicable to the problem considered in this
paper. One reason is that the sparse representation techniques developed
in \cite{Shahmansoori} and \cite{Re2LilongDai}, based on the DFT
of the array steering vectors, are only suitable for ULAs and URAs,
where linear recurrence relations exist between the array steering
vectors; see \cite[eq. (5)]{Shahmansoori} and \cite[eq. (3)]{Re2LilongDai}.
Another reason is that the method developed in \cite{Re2LilongDai}
only estimates the so-called channel component which is the product
of channel parameters (including the channel gain and signal direction),
and does not estimate explicitly the channel parameters, e.g., the
angle and delay. The method in \cite{Shahmansoori} uses the ML-based
algorithms to estimate the channel parameters (i.e., angle and delay)
in the same way as Q-MLE \cite{QMLE}, which is one of the benchmarks
used in our performance evaluation.

As shown in Figs. \ref{fig:result0}(a) and \ref{fig:result0}(b),
the proposed method and GBM are worse than QRIPM in terms of angle
estimation, when the number of antennas is small, i.e., less than
100. The reason is that the proposed algorithm may suffer from an
inaccurate approximation in \eqref{eq:AHm}, due to the unsatisfied
conditions in Theorem 1. However, as the number of antennas increases,
the accuracies of the proposed approach and GBM improve faster than
that of QRIPM. The proposed method quickly outperforms
both QRIPM and GBM, and approaches the CRLB. The improvement slows
down with the increasing number of antennas. When the number of antennas
is large (e.g., more than 300), the estimation accuracies of the AOAs
improve marginally, resulting from the increasingly negligible relative
growth of the array aperture of the circular arrays. As also shown
in Figs. \ref{fig:result0}(a) and \ref{fig:result0}(b), Q-MLE outperforms
the other three approaches, including the proposed approach, in terms
of angle estimation. However, Q-MLE has a significantly higher computational
complexity than the proposed approach, as discussed in Section \ref{subsec:Complexity-Analysis}.

As shown in Fig. \ref{fig:result0}(c), the proposed
approach achieves the best delay estimation accuracy, attributing
to the high temporal resolution of the wideband mmWave signals offered
by the MDSI method in the proposed approach. The RMSE curves of the
estimated delay appear to be constant. The reason is because the delay
estimation precision depends primarily on the signal bandwidth, and
is less affected by the number of antennas at the BS (as opposed to
the angle estimation). Given its superiority in
the angle and delay estimation, the proposed approach outperforms
QRIPM, GBM, and Q-MLE in terms of localization, as corroborated in
Fig. \ref{fig:result0}(d).

\begin{figure*}
\begin{centering}
\includegraphics[width=14cm]{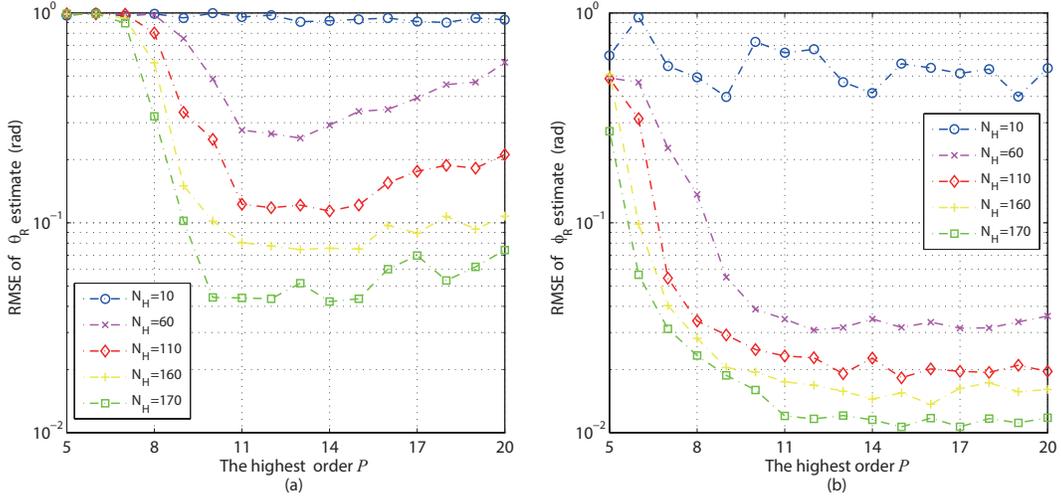}
\par\end{centering}
\centering{}\caption{The RMSE vs. the value of the highest order $P$. (a) The azimuth
AOA; and (b) The elevation AOA. \label{fig:Pp}}
\end{figure*}

In order to validate Theorem 1, Fig. \ref{fig:Pp} plots the RMSE
of the angle estimation versus the value of the highest order, $P$,
under different numbers of horizontal array response vectors. We see
that when the highest order $P\leq11$, our proposed approach cannot
perform satisfactorily, since the number of phase-mode vectors is
not sufficient to represent the transformed array response vectors
in Section \ref{subsec:Step-2:-Horizontal}. Fig. \ref{fig:Pp} also
shows that, if $P\geq12$, for any number of array response vectors,
increasing the phase-mode vectors has little influence on the angle
estimation performance. This means that the number of phase-mode vectors
needed in our approach does not depend on the number of array response
vectors, which is important for complexity reduction, as discussed
in Section \ref{subsec:Complexity-Analysis}. In addition, we also
see that because the condition in Theorem 1, $N_{\mathrm{H}}\geq2P$,
is unlikely to be satisfied when $N_{\mathrm{H}}=10$, the RMSE is
much poorer than those applying more array response vectors.

\begin{figure*}
\begin{centering}
\includegraphics[width=14cm]{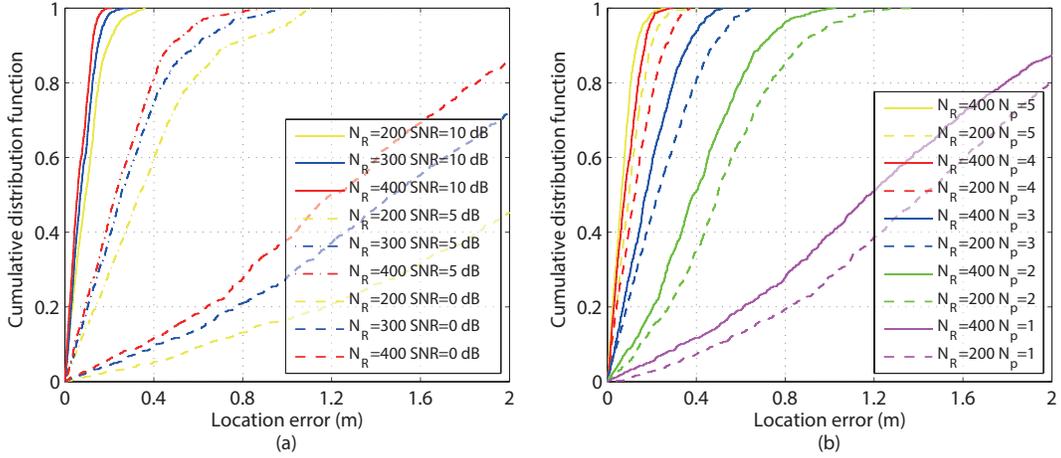}
\par\end{centering}
\centering{}\caption{The CDF of the localization error for different numbers of receive
antennas in the case of (a) different received SNR values; and (b)
different numbers of received paths\label{fig:CDF}}
\end{figure*}

Fig. \ref{fig:CDF} assesses statistically the proposed approach by
plotting the cumulative distribution function (CDF) of the localization
error. We assume that the time offset $\tau_{\textrm{of}}$ obeys
a zero-mean Gaussian distribution with the standard deviation of 4
ns, and $\tau_{\textrm{of}}$ is unknown to the BS in the simulation.
In Fig. \ref{fig:CDF}(a), we observe that although the performance
of the proposed approach decreases with the decline of the average
received SNR, the statistical localization error remains small even
for SNR = 0 dB, as long as a sufficient number of receive antennas
are deployed. In addition, we see that the proposed approach is able
to achieve a centimeter-level localization accuracy with a probability
of over 60\%, when the number of receive antennas is 200. Fig. \ref{fig:CDF}(b)
shows the relationship between the localization accuracy and the number
of received paths. It can be seen that the proposed approach cannot
provide high-accuracy localization with high probability if only a
single path is received, since the time offset $\tau_{\textrm{of}}$
is unknown to the BS, as discussed in Section \ref{subsec:Positioning-Method}.
In Fig. \ref{fig:CDF}(b), we also see that when the number of received
paths is more than three, more paths lead to limited improvement in
localization accuracy.

\section{Conclusion \label{sec:Conclusion}}

In this paper, a novel joint delay and angle estimation approach was
proposed for wideband mmWave large-scale hybrid arrays. We proposed
a new 3D hybrid beamformer to reduce the number of required RF chains
while maintaining the critical recursive property of the space-time
response matrix for angle and delay estimation. We also generalized
linear interpolation to reconstruct the output signals of the 3D hybrid
beamformer and to achieve consistent array response across the wideband
and suppress the beam squint effect. As a result, the delay and the
azimuth and elevation angles of every multi-path component can be
estimated. Simulation results showed that, when a large number of
antennas is deployed, our proposed approach is capable of precisely
estimating the channel parameters even in low SNR regimes. Potential
future extensions of this work include simultaneous localization and
environment mapping, performance evaluation of the proposed approach
in real-world scenarios, and new techniques to accelerate parameter
matching.

\section*{Appendix I\protect \protect \protect \protect \protect \protect
\protect \protect \protect \protect \protect \\
 Proof of Lemma 1}

According to the property of Bessel function, i.e., $J_{-v}(x)=(-1)^{v}J_{v}(x)$,
we have $\left|J_{-v}(x)\right|=\left|J_{v}(x)\right|$, so here we
only use $J_{v}(x)$ with $v\in\mathbb{Z}^{+}$ for illustration convenience.
Let $x=v\rho,\rho\in(0,1]$. The Bessel function, $J_{v}(x)$, whose
order $v$ exceeds its argument, $x$, can be written in the following
form \cite{bessel}

\begin{equation}
J_{v}(v\rho)=\frac{1}{\pi}\int_{0}^{\pi}\exp\left(-vF(\vartheta,\rho)\right)d\vartheta,\label{eq:bessel}
\end{equation}
where
\begin{align*}
F(\vartheta,\rho) & =\log\left(\frac{\vartheta+\sqrt{\vartheta^{2}-\rho^{2}\sin^{2}\vartheta}}{\rho\sin\vartheta}\right)\\
 & -\cot\vartheta\sqrt{\vartheta^{2}-\rho^{2}\sin^{2}\vartheta}.
\end{align*}

The partial derivative of \eqref{eq:bessel} with respect to $\rho$
is calculated as

\begin{align}
\frac{\partial}{\partial\rho}J_{v}(v\rho)=-\frac{v}{\pi}\int_{0}^{\pi}\frac{\partial F(\vartheta,\rho)}{\partial\rho}\exp\left(-vF(\vartheta,\rho)\right)d\vartheta\nonumber \\
=\frac{v}{\pi\rho}\int_{0}^{\pi}g(\vartheta,\rho)\exp\left(-vF(\vartheta,\rho)\right)d\vartheta,\label{eq:besselpartial}
\end{align}
where $g(\vartheta,\rho)=\left(\vartheta-\rho^{2}\sin\vartheta\cos\vartheta\right)/\sqrt{\vartheta^{2}-\rho^{2}\sin^{2}\vartheta}.$
Considering that
\begin{align}
g(\vartheta,\rho) & =\frac{\vartheta-\rho^{2}\sin\vartheta\cos\vartheta}{\sqrt{\vartheta^{2}-\rho^{2}\sin^{2}\vartheta}}\geq\frac{\vartheta-\sin\vartheta\cos\vartheta}{\sqrt{\vartheta^{2}-\rho^{2}\sin^{2}\vartheta}}\nonumber \\
 & \geq\frac{\vartheta-\sin\vartheta}{\sqrt{\vartheta^{2}-\rho^{2}\sin^{2}\vartheta}}\geq0,
\end{align}
we have $\partial J_{v}(v\rho)/\partial\rho>0,$ and conclude that
$J_{v}(v\rho)$ is a positive increasing function of $\rho$. Thus,
$J_{v}(v\rho)<J_{v}(v).$

On the other hand, the partial derivative of \eqref{eq:bessel} with
respect to $v$ is calculated as
\begin{equation}
\frac{\partial}{\partial v}J_{v}(v\rho)=-\frac{1}{\pi}\int_{0}^{\pi}F(\vartheta,\rho)\exp\left(-vF(\vartheta,\rho)\right)d\vartheta.
\end{equation}
Because
\begin{equation}
\frac{\partial}{\partial\vartheta}F(\vartheta,\rho)=\frac{(1-\rho\cot\vartheta)^{2}}{\sqrt{\vartheta^{2}-\rho^{2}\sin^{2}\vartheta}}+\sqrt{\vartheta^{2}-\rho^{2}\sin^{2}\vartheta}\geq0
\end{equation}
and $\partial F(0,\rho)/\partial\rho=-\sqrt{1-\rho^{2}}/\rho\leq0$,
we have $F(\vartheta,\rho)\geq F(0,\rho)\geq F(0,1)=0$, and hence
$\partial J_{v}(v\rho)/\partial v<0$. This means that $J_{v}(v\rho)$
is a positive decreasing function of $v$, i.e., $J_{v}(v\rho)\leq J_{1}(\rho).$
Therefore, we have $J_{v}(v\rho)<J_{v}(v)\leq J_{1}(1)\approx0.44$
with $\rho\in(0,1]$ and $v\in\mathbb{Z}^{+}$. When $\left|v\right|>\left|x\right|$,
$\left|J_{v}(x)\right|\approx0,\:v\in\mathbb{Z}.$

\section*{Appendix II\protect \protect \protect \protect \protect \protect
\protect \protect \protect \protect \protect \\
 Proof of Theorem 1}

According to Lemma 1, we observe that $J_{p}(\varpi_{m,l})$ cannot
be omitted if $\left|p\right|\leq\left|\varpi_{m,l}\right|=\left|2\pi f_{m}r\sin(\theta_{\textrm{R},l})/c\right|\leq2\pi f_{m}r/c$.
Because $f_{0}\leq f_{m}$ and $p\in\mathbb{Z}$, we set the highest
order $P=\max(\left|p\right|)=\left\lfloor 2\pi f_{0}r/c\right\rfloor $.

On the other hand, in the case of $Q\neq0$, because $p\in[-P,P]\cap\mathbb{Z}$
and $N_{\mathrm{H}}\geq2P$, we have $\left|p-QN_{\mathrm{H}}\right|\geq\left|\varpi_{m,l}\right|$.
According to Lemma 1, we obtain
\begin{align}
 & \left|\varepsilon_{p,Q}(\varpi_{m,l},\phi_{\textrm{R},l})\right|\nonumber \\
 & =\left|j^{(QN_{\mathrm{H}}-p)}J_{(QN_{\mathrm{H}}-p)}(\varpi_{m,l})\exp\left(j(QN_{\mathrm{H}}-p)\phi_{\textrm{R},l}\right)\right|\nonumber \\
 & =\left|J_{(p-QN_{\mathrm{H}})}(\varpi_{m,l})\right|\approx0.
\end{align}
In this case, \eqref{eq:dft2} can be approximated by
\begin{align}
A_{\textrm{PM},p} & =\sqrt{N_{\textrm{H}}}\left[j^{p}J_{p}(\varpi_{m,l})e^{-jp\phi_{\textrm{R},l}}\vphantom{\sum_{Q\neq0}^{\infty}}\right.\nonumber \\
 & \left.+\sum_{Q=-\infty,Q\neq0}^{\infty}\varepsilon_{p,Q}(\varpi_{m,l},\phi_{\textrm{R},l})\right]\nonumber \\
 & \approx\sqrt{N_{\textrm{H}}}j^{p}J_{p}(\varpi_{m,l})e^{-jp\phi_{\textrm{R},l}}.
\end{align}
This concludes the proof of Theorem 1.

\begin{IEEEbiography}[{\includegraphics[width=1in,height=1.25in,clip,keepaspectratio]{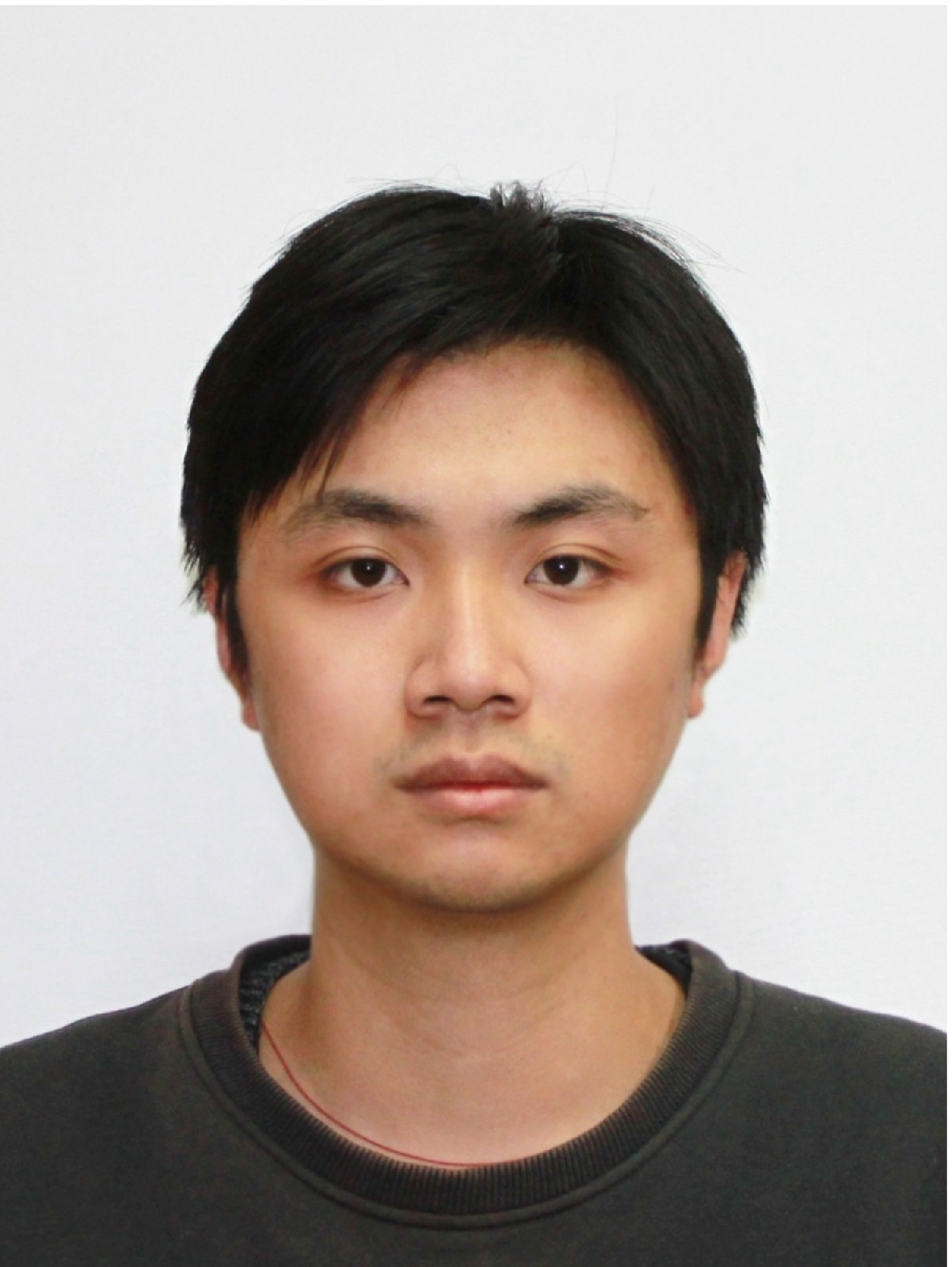}}]{Zhipeng Lin}

(M'20)  is currently working toward the dual Ph.D. degrees  in communication and information engineering in the School of Information and Communication Engineering, Beijing University of Posts and Telecommunications, Beijing, China, and the School of Electrical and Data Engineering, University of Technology of Sydney,  NSW, Australia. His current research interests include millimeter-wave communication, massive MIMO, hybrid beamforming, wireless localization, and tensor processing.

\end{IEEEbiography}

\begin{IEEEbiography}[{\includegraphics[width=1in,height=1.25in,clip,keepaspectratio]{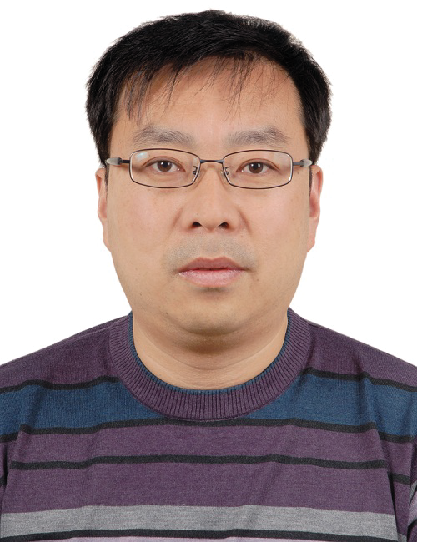}}]{Tiejun Lv}
(M'08-SM'12) received the M.S. and Ph.D. degrees in electronic engineering from the University of Electronic Science and Technology of China (UESTC), Chengdu, China, in 1997 and 2000, respectively. From January 2001 to January 2003, he was a Postdoctoral Fellow with Tsinghua University, Beijing, China. In 2005, he was promoted to a Full Professor with the School of Information and Communication Engineering, Beijing University of Posts and Telecommunications (BUPT). From September 2008 to March 2009, he was a Visiting Professor with the Department of Electrical Engineering, Stanford University, Stanford, CA, USA. He is the author of 3 books, more than 80 published IEEE journal papers and 190 conference papers on the physical layer of wireless mobile communications. His current research interests include signal processing, communications theory and networking. He was the recipient of the Program for New Century Excellent Talents in University Award from the Ministry of Education, China, in 2006. He received the Nature Science Award in the Ministry of Education of China for the hierarchical cooperative communication theory and technologies in 2015.
\end{IEEEbiography}

\begin{IEEEbiography}[{\includegraphics[width=1in,height =1.25in,clip,keepaspectratio]{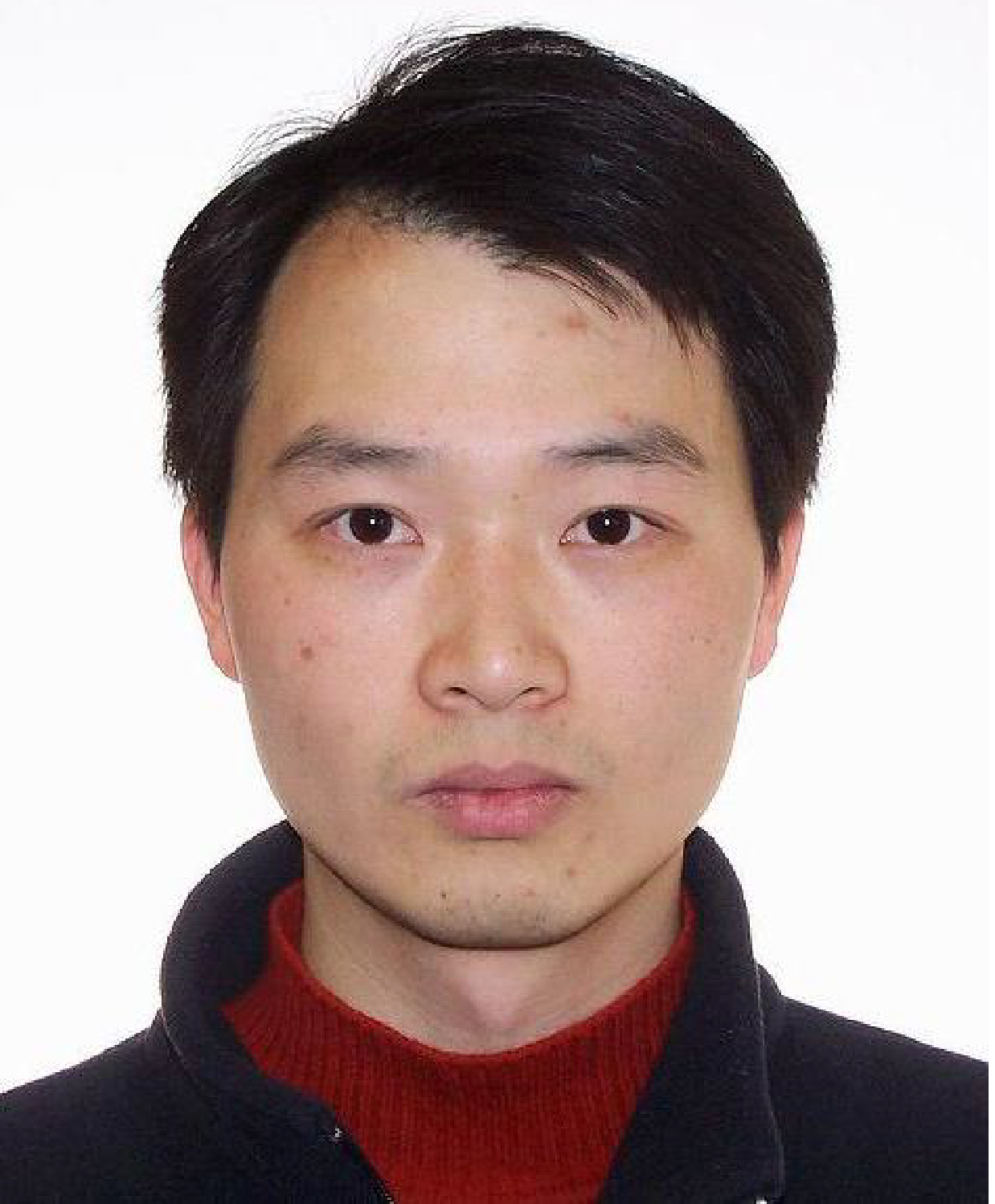}}]{Wei Ni}
(M'09-SM'15) received the B.E. and Ph.D. degrees in Electronic Engineering from Fudan University, Shanghai, China, in 2000 and 2005, respectively. Currently, he is a Group Leader and Principal Research Scientist at CSIRO, Sydney, Australia, and an Adjunct Professor at the University of Technology Sydney and Honorary Professor at Macquarie University, Sydney. He was a Postdoctoral Research Fellow at Shanghai Jiaotong University from 2005 -- 2008; Deputy Project Manager at the Bell Labs, Alcatel/Alcatel-Lucent from 2005 to 2008; and Senior Researcher at Devices R\&D, Nokia from 2008 to 2009. His research interests include signal processing, stochastic optimization, learning, as well as their applications to network efficiency and integrity.

Dr Ni is the Chair of IEEE Vehicular Technology Society (VTS) New South Wales (NSW) Chapter since 2020 and an Editor of IEEE Transactions on Wireless Communications since 2018. He served first the Secretary and then Vice-Chair of IEEE NSW VTS Chapter from  2015 to 2019, Track Chair for VTC-Spring 2017, Track Co-chair for IEEE VTC-Spring 2016, Publication Chair for BodyNet 2015, and  Student Travel Grant Chair for WPMC 2014.
\end{IEEEbiography}

\begin{IEEEbiography}[{\includegraphics[width=1in,height=1.25in,clip,keepaspectratio]{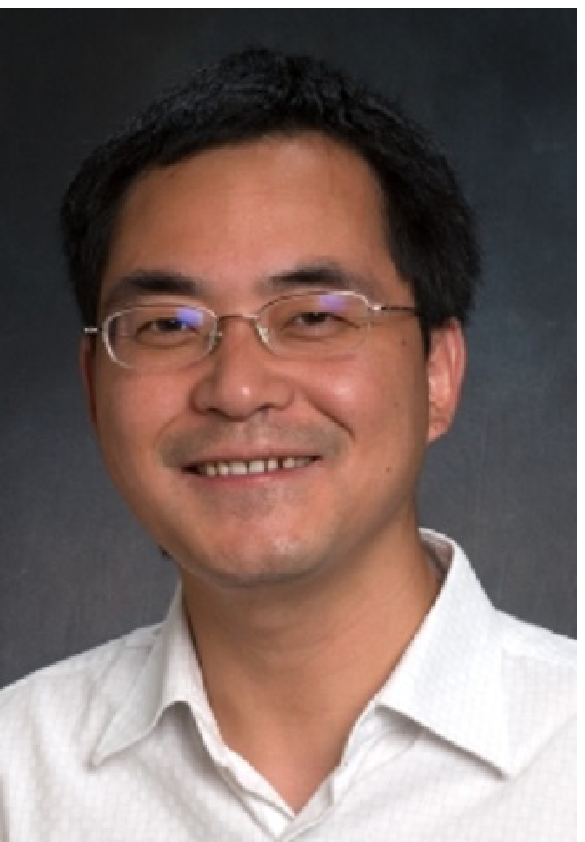}}]{J. Andrew Zhang}
(M'04-SM'11) received the B.Sc. degree from Xi'an JiaoTong University, China, in 1996, the M.Sc. degree from Nanjing University of Posts and Telecommunications, China, in 1999, and the Ph.D. degree from the Australian National University, in 2004.

Currently, Dr. Zhang is an Associate Professor in the School of Electrical and Data Engineering, University of Technology Sydney, Australia. He was a researcher with Data61, CSIRO, Australia from 2010 to 2016, the Networked Systems, NICTA, Australia from 2004 to 2010, and ZTE Corp., Nanjing, China from 1999 to 2001.  Dr. Zhang's research interests are in the area of signal processing for wireless communications and sensing. He has published more than 180 papers in leading international Journals and conference proceedings, and has won 5 best paper awards. He is a recipient of CSIRO Chairman's Medal and the Australian Engineering Innovation Award in 2012 for exceptional research achievements in multi-gigabit wireless communications.

\end{IEEEbiography}

\begin{IEEEbiography}[{\includegraphics[width=1in,height=1.25in,clip,keepaspectratio]{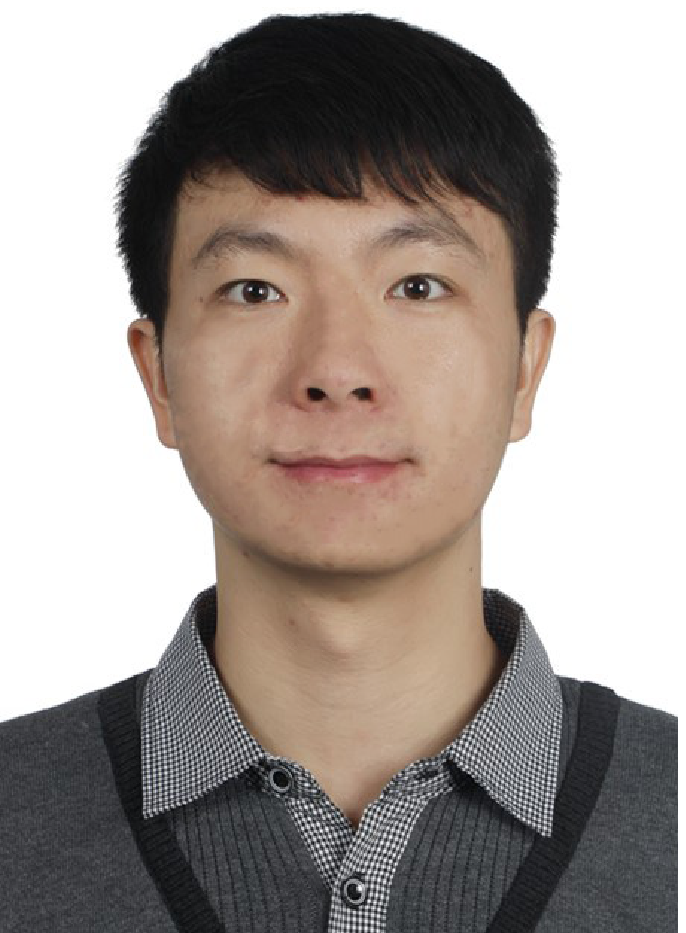}}]{Jie Zeng}
(M'09--SM'16)  received the B.S. and M.S. degrees from Tsinghua University in 2006 and 2009, respectively, and received the Ph.D. degree from Beijing University of Posts and Telecommunications in 2019. From 2009, he has been with Tsinghua University. His research interests include 5G, IoT, URLLC, novel multiple access, and novel network architecture. He has authored three books related to 5G, has published over 100 journal and conference papers, and holds more than 30 Chinese and international patents. He participated in drafting one national standard and one communication industry standard in China. He received the science and technology award of Beijing in 2015 and the best cooperation award of Samsung Electronics in 2016.

\end{IEEEbiography}

\begin{IEEEbiography}[{\includegraphics[width=1in,height=1.25in,clip,keepaspectratio]{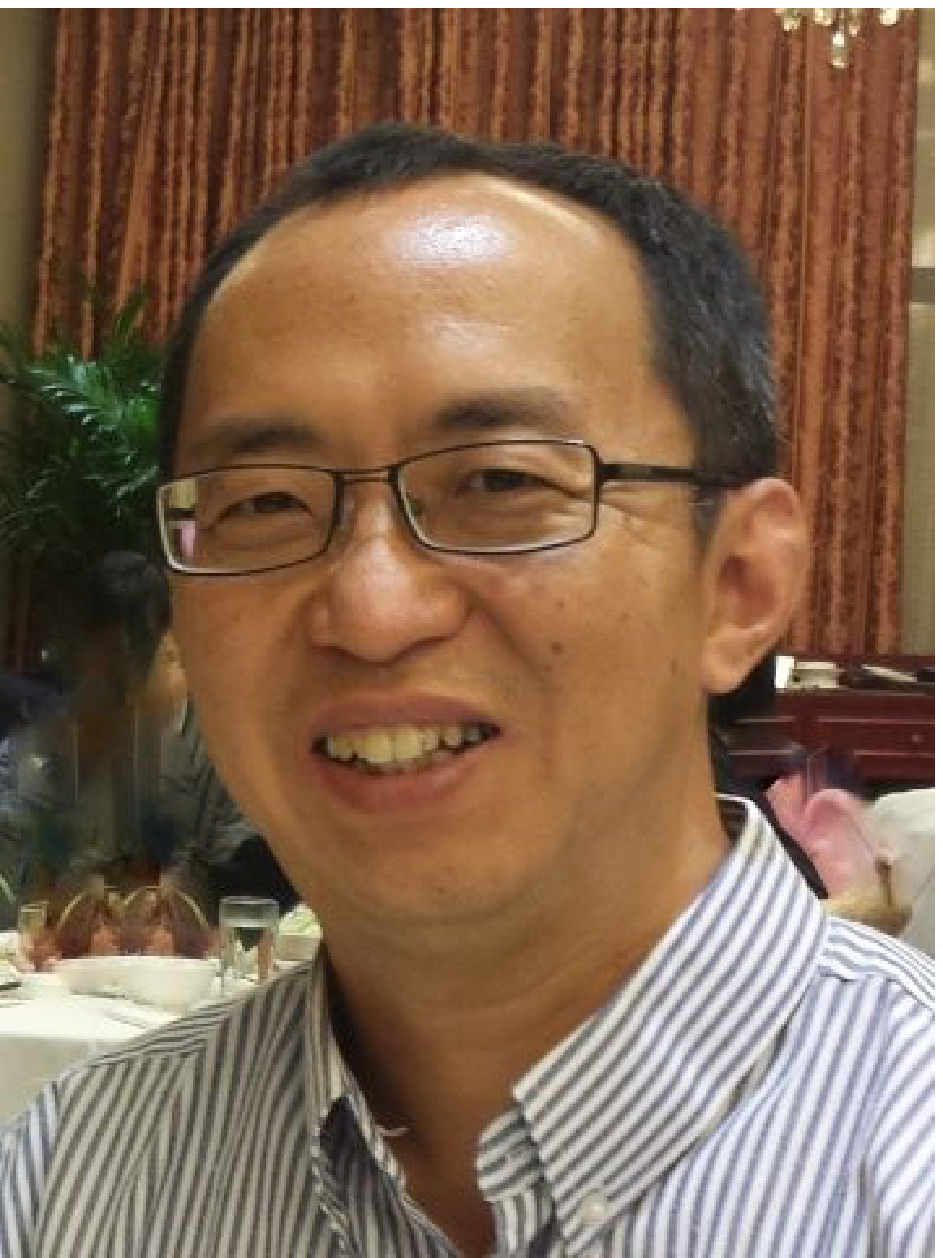}}]{Ren Ping Liu}
(M'09-SM'14) received his B.E. and M.E. degrees from Beijing University of Posts and Telecommunications, China, and the Ph.D. degree from the University of Newcastle, Australia.

He is currently a Professor and Head of Discipline of Network \& Cybersecurity at University of Technology Sydney. Professor Liu was the co-founder and CTO of Ultimo Digital Technologies Pty Ltd, developing IoT and Blockchain. Prior to that he was a Principal Scientist and Research Leader at CSIRO, where he led wireless networking research activities. He specialises in system design and modelling and has delivered networking solutions to a number of government agencies and industry customers. His research interests include wireless networking, Cybersecurity, and Blockchain.

Professor Liu was the founding chair of IEEE NSW VTS Chapter and a Senior Member of IEEE. He served as Technical Program Committee chairs and Organising Committee chairs in a number of IEEE Conferences. Prof Liu was the winner of Australian Engineering Innovation Award and CSIRO Chairman medal. He has over 200 research publications.

\end{IEEEbiography}

\end{document}